\documentclass[a4paper,onecolumn,10pt]{quantumarticle}
\pdfoutput=1
\usepackage[utf8]{inputenc}
\usepackage[english]{babel}
\usepackage[T1]{fontenc}
\usepackage{amsmath}
\usepackage{amsfonts}
\usepackage{hyperref}
\usepackage{braket}
\usepackage{quantikz}
\usepackage{lipsum}

\usepackage{xcolor}
\usepackage[normalem]{ulem}
\usepackage{afterpage}

  \usepackage[backend=biber, style=numeric-comp, maxbibnames=1, minbibnames=1, sorting=none]{biblatex}
 \AtBeginBibliography{\sloppy}
\begin{document}

\title{Quantum Search without Global Diffusion}

\author{John Burke}
\email{burkej15@tcd.ie}
\affiliation{School of Computer Science and Statistics, Trinity College Dublin, College Green, Dublin 2, Ireland}
\author{Ciaran Mc Goldrick}
\email{ciaran.mcgoldrick@tcd.ie}
\affiliation{School of Computer Science and Statistics, Trinity College Dublin, College Green, Dublin 2, Ireland}
\maketitle

\begin{abstract}
Quantum search is among the most important algorithms in quantum computing, offering fundamentally new approaches to optimisation and information processing. At its core is quantum amplitude amplification, which achieves a quadratic speedup over classical search approaches. This technique alternates applications of two reflections: the oracle, which marks the targets, and the global diffusion operator, which reflects about the initial state and acts on every qubit that encodes the search space. We show that the quadratic speedup can be preserved for search problems such as single-target unstructured search, even when the operations between any two successive oracle calls are restricted to a small subset of these qubits. We present a recursive construction that, when the state preparation unitary and the projector onto the marked subspace factorise across a partition of the search register, acts on only one register at a time and admits an exact solution for the algorithm's dynamics. This is enabled by an intriguing degeneracy in the principal angles between successive reflections, which collapse to at most two distinct values governed by a single recursively defined angle.

Applied to single-target unstructured search, which naturally satisfies this factorisation, the approach retains the $O(\sqrt{N})$ oracle complexity of Grover search while acting on as few as $\log_2(\log_2 N)$ qubits between successive oracle calls. On an 18-qubit search problem, partitioning into three stages reduces the expected non-oracle circuit depth by as much as $50\%$--$86\%$ relative to Grover, depending on the gate decomposition used, requiring $6.7\%$ additional expected oracle calls. For larger problem sizes with sufficiently large partitions, this oracle overhead rapidly diminishes, and valuable depth reductions persist when the oracle circuit is substantially deeper than the diffusion operator. More broadly, these results show that quantum search does not need to manipulate more than a small subset of qubits between successive oracle calls to achieve its quadratic speedup, offering a new perspective on this foundational algorithm. The scalar reduction at the heart of our analysis also motivates new directions in quantum algorithm design and evaluation.

\end{abstract}
\section{Introduction}

Quantum amplitude amplification~\cite{AmplitudeAmplification} remains one of the most widely employed primitives in quantum computing, generalising the quadratic speedup of Grover's search algorithm~\cite{Grover} to a broader class of problems. Given a quantum algorithm $A$ that produces a state with some overlap with the target, amplitude amplification boosts the probability of measuring the target using $O(1/\sqrt{p})$ applications of $A$ and its inverse, where $p$ is the success probability when measuring the initial state. For unstructured search, this quadratic speedup has been shown to be asymptotically optimal~\cite{Optimal} and has found applications in domains such as combinatorial optimisation~\cite{GroverNP}, classification~\cite{groverImage}, and machine learning~\cite{groverMachineLearning}.

At the core of amplitude amplification are two global reflections: the phase oracle $S_x$ and the diffusion operator $S_\psi$. The oracle reflects about the target state $\ket{x}$, while the diffusion operator reflects about a starting state $\ket{\psi} = A\ket{0}$ prepared by a quantum algorithm $A$. Their product is a rotation on the two-dimensional plane spanned by these states, a conceptually elegant structure that underpins the quadratic reduction in query complexity. Although this query complexity is provably optimal~\cite{Optimal}, significant research has extended the algorithm in other promising directions, including fixed-point amplification~\cite{FixedPointSearch, fixed-point}, oblivious amplification~\cite{OAA}, deterministic variants~\cite{AbitraryPhases, GroverZeroFailRate, DeterministicRestrictedOracle, DeterministicAdjustableParameters}, and alterations to the non-oracle operations to make the approach more practical to implement~\cite{grover_trade-offs_2002, searchDepthOptimization, hardware, groverOptGates, hardwareGrover2}.

Central to the latter line of work is the \emph{partial} diffusion operator, which has the same reflective structure as the global diffuser, but instead acts on a subset of the search register. Unlike the global diffusion operator, which acts on every qubit of the search register and reflects about a single state vector, the partial diffusion operator reflects about a subspace, and can be used to amplify a component of the target state. In practice, this operator requires the quantum algorithm $A$ to decompose as a tensor product over the search space. Such a condition is naturally satisfied by Grover search, where the initial uniform superposition can be created by applying a Hadamard operation to each individual qubit. This operator is integral to partial quantum search~\cite{partialSearch, partialSearch2}, where only part of the target state is required, and can offer a slight improvement in oracle complexity over full quantum search. The partial diffusion operator has since been applied to full search~\cite{grover_trade-offs_2002, searchDepthOptimization, hardware, groverOptGates, hardwareGrover2, zhang2026palindromic} as a means of reducing the non-oracle operator cost for implementation.

While partial diffusers substantially reduce the number of global diffusion operators required for quantum search, their use introduces significant analytical challenges. Most fundamentally, they break the two-dimensional rotational geometry that makes standard amplitude amplification so tractable. Alternating between partial and global diffusion operators --- the structure underpinning partial search algorithms --- gives rise to rotations within a three-dimensional subspace~\cite{so3}. Replacing the global operator entirely is harder still and the dynamics are governed by a unitary acting on a subspace that grows exponentially with the number of distinct diffusion operators applied. Even in the simplest non-trivial case of two operators, computing the success probability after $t$ iterations requires evaluating the $t$-th power of a $4\times 4$ unitary~\cite{hardware}. Alternative approaches avoid this complexity by using partial diffusion operators to construct intermediate algorithms that concentrate amplitude towards the target value by a known amount, then feeding the resulting circuit into standard amplitude amplification as the algorithm $A$~\cite{grover_trade-offs_2002, groverOptGates, hardwareGrover2}.

These methods, however, only reduce the number of global diffusion operators rather than eliminating them. While these constructions largely act on small subsets of qubits between oracle calls, reaching as few as $O(\sqrt{N})$ non-oracle gates in total~\cite{hardwareGrover2}, the amplitude amplification wrapper still applies an $n$-qubit reflection. Other constructions replace the reflection about $\ket{\psi}$ with a layer of single-qubit gates~\cite{kato2005single, jiang2017transverse}. However, this layer still acts on every qubit between oracle calls, and these constructions need a phase-tuned oracle~\cite{kato2005single} or succeed with probability about one half~\cite{jiang2017transverse}. The method of~\cite{hardware} splits the search register into two parts, only one of which is acted on between successive oracle calls. One part, however, always holds at least $n/2$ qubits, and the query count is established only numerically. Without the wrapper amplitude amplification, two-qubit diffusion operators alone still solve single-target search exactly, as noted in~\cite{hardwareGrover2} and developed in our earlier work~\cite{burke2025recursive}, but the number of oracle calls grows to $O(N^{(\log_2 3)/2})$, with $(\log_2 3)/2\approx 0.79$, and the quadratic speedup is lost. Table~\ref{tab:comparison} compares these approaches.
\begin{table}[t]
\centering
\footnotesize
\setlength{\tabcolsep}{4pt}
\begin{tabular}{|l|l|l|p{4.1cm}|}
\hline
Work & Oracle calls & Non-oracle gates & Qubits acted on between oracle calls (worst / average) \\
\hline
Grover~\cite{Grover} & $O(\sqrt{N})$ & $\Theta(\sqrt{N}\log N)$ & $n$ / $n$ \\
\hline
Grover~\cite{grover_trade-offs_2002} & $O(\sqrt{N})$ & $O(\sqrt{N}\log\log N)$ & $n$ / $O(\log n)$ \\
\hline
Arunachalam and de Wolf~\cite{groverOptGates} & $O(\sqrt{N})$ & $O(\sqrt{N}\log^{(r)} N)$ & $n$ / $O(\log^{(r)} N)$ \\
\hline
Bria\'nski et al.~\cite{hardwareGrover2} & $O(\sqrt{N})$ & $\Theta(\sqrt{N})$ & $n$ / $O(1)$ \\
\hline
Zhang and Korepin~\cite{searchDepthOptimization, zhang2026palindromic} & $O(\sqrt{N})$ & $\Theta(\sqrt{N}\log N)$ & $n$ / $\Theta(n)$ \\
\hline
Kato~\cite{kato2005single} & $O(\sqrt{N})$, phase-$\omega$ & $\Theta(\sqrt{N}\log N)$ & $n$ / $n$ (single-qubit gates) \\
\hline
Jiang et al.~\cite{jiang2017transverse} & $O(\sqrt{N})$, success $\approx 1/2$ & $\Theta(\sqrt{N}\log N)$ & $n$ / $n$ (single-qubit gates) \\
\hline
Liu and Zhou~\cite{hardware} & $O(\sqrt{N})$, numerical & $\Theta(\sqrt{N}\log N)$ & $\Theta(n)$ / $\Theta(n)$ \\
\hline
Two-qubit case~\cite{hardwareGrover2, burke2025recursive} & $O(N^{(\log_2 3)/2})$ & $O(N^{(\log_2 3)/2})$ & $2$ / $2$ \\
\hline
This work & $O(\sqrt{N})$ & $O(\sqrt{N}\log\log N)$ & $\Theta(\log n)$ / $\Theta(\log n)$ \\
\hline
\end{tabular}
\caption{Comparison of single-target quantum search algorithms on $n$ qubits, with $N=2^n$. The algorithms in~\cite{groverOptGates, hardwareGrover2, burke2025recursive} are deterministic, those in~\cite{Grover, grover_trade-offs_2002, searchDepthOptimization, zhang2026palindromic, kato2005single, hardware} and this work have a success probability that tends to one, and that in~\cite{jiang2017transverse} has a success probability of about one half. The last column gives the largest number of qubits acted on between two successive oracle calls, and the average over all such pairs. For each construction, gate counts assume that a diffusion operator has a cost linear in the number of qubits it acts on. The algorithm of~\cite{hardwareGrover2} uses at most $(1+\epsilon)(\pi/4)\sqrt{N}$ oracle calls and $O(\sqrt{N}\log(1/\epsilon))$ non-oracle gates, both $O(\sqrt{N})$ for a fixed $\epsilon$. The construction of~\cite{kato2005single} uses a phase-$\omega$ oracle in place of the standard reflection, the oracle and gate counts of~\cite{hardware} are established numerically, and $\log^{(r)}$ denotes the logarithm applied $r$ times.}
\label{tab:comparison}
\end{table}

In this work we demonstrate that the quadratic speedup can be achieved for search problems such as single-target unstructured search, even when only a small subset of qubits is operated on between any two successive oracle calls. Specifically, when the state preparation unitary and the projector onto the marked subspace factorise across a fixed partition of the search register, we can construct a recursive search procedure using partial diffusion operators on the individual registers of this partition. This procedure admits an exact solution for its dynamics, and for single intermediate iterations we derive an explicit bound on its oracle complexity. Each iteration of the algorithm concentrates amplitude within one register onto the corresponding section of the target state. The remaining registers are then processed in turn with a sequence of inverse search iterates, with a single measurement at the end. For single-target unstructured search, partitioning the search register into two subsets and using a single intermediate iteration recovers the circuit of~\cite{hardware}, whose success probability and query count we thereby obtain analytically. We show that while the operator eigenspaces grow exponentially with the number of partitions, the principal angles between successive reflection operators collapse to at most two distinct values. These are governed by a single recursively defined angle determined by the overlap between the initial and target states within each register. After $m$ layers of recursion, the action of the algorithm reduces to two distinct rotations whose projection onto the target component is determined entirely by this angle and the outer iteration count. The oracle overhead relative to standard quantum amplitude amplification is $O(1/\prod_{i=1}^{m}\cos(\theta_i))$, where $\sin^2(\theta_i) = p_i$ is the initial success probability when measuring register $i$. For $0<p_i<1/2$, this overhead is $O(1)$ whenever $\sum_{i=1}^{m}p_i=O(1)$. 
Applied to unstructured search of a single target, a problem that naturally satisfies the tensor product decomposition, the construction retains the $O(\sqrt{N})$ query complexity when each partition contains at least $\log_2(\log_2 N)$ qubits. For an 18-qubit search problem partitioned into three stages, the expected non-oracle circuit depth can be reduced by $50\%$--$86\%$ relative to Grover, depending on the gate decomposition used, at a cost of $6.7\%$ additional expected oracle calls. For larger problem sizes, we show that total circuit depth savings persist for stages of at least $\sqrt{n}$ qubits, even when the circuit depth of the oracle is substantially deeper than that of the diffusion operator. These depth reductions lower both execution time and noise susceptibility, a practical benefit at any scale. More broadly, this work demonstrates that the quadratic speedup of quantum search can be achieved when the operations between any two successive oracle calls act only on a small subset of qubits, with the oracle as the only operator acting across more than one such subset. The scalar reduction at the heart of this result offers new perspectives on both the algorithm itself and the design of future quantum algorithms.

\section{Preliminaries}

\subsection{Quantum Amplitude Amplification}
\label{subsec:qaa}

Quantum amplitude amplification~\cite{AmplitudeAmplification} generalises the quadratic speedup of Grover's search algorithm to a wider range of optimisation tasks beyond unstructured search. Given a quantum algorithm $A$ that prepares an initial state $\ket{\psi} = A\ket{0}$ and a set of orthonormal target states $\{\ket{z_1},\ldots,\ket{z_r}\}$, the aim of amplitude amplification is to transform $\ket{\psi}$ into an output state $\ket{x}$ that lies entirely within the subspace spanned by these targets. Measuring $\ket{x}$ in a basis containing the target states would then yield one of the targets with unit probability. Let $P_G$ denote the orthogonal projector onto this marked subspace. The initial probability of success is the squared norm of the projection of $\ket{\psi}$ onto this subspace, $p = \|P_G\ket{\psi}\|^2$, corresponding to the probability of obtaining a target outcome when measuring $\ket{\psi}$ in the same basis. For $0 < p < 1$, we define the desired output state $\ket{x} = P_G\ket{\psi}/\sqrt{p}$ as the normalised projection of the initial state onto the marked subspace, with $\braket{x|\psi} = \sqrt{p}$. Amplitude amplification then transforms $\ket{\psi}$ into a state with high overlap with $\ket{x}$ in $O(1/\sqrt{p})$ steps. The amplification is realised by repeatedly applying the iterate
\begin{equation}
    Q = S_\psi \, S_x
\end{equation}
where $S_x = \mathbb{I}_n - 2P_G$ is the oracle, a reflection that applies a phase of $-1$ to the marked subspace and leaves its orthogonal complement unchanged, and $S_\psi = \mathbb{I}_n - 2\ket{\psi}\bra{\psi} = A(\mathbb{I}_n - 2\ket{0}\bra{0})A^\dagger$ is a reflection about the initial state, also known as the diffusion operator. The iterate $Q$ is therefore a product of two reflections. As per Jordan's lemma, any such product can be simultaneously block-diagonalised into $1\times 1$ and $2\times 2$ blocks, where the $2\times 2$ blocks correspond to invariant rotation planes characterised by a principal angle between the reflected eigenspaces. The plane $\mathcal{K} = \mathrm{span}\{\ket{\psi},\ket{x}\}$, spanned by the initial and desired output states, is invariant under both reflections. Within this plane, $P_G$ acts as the rank-one projector $\ket{x}\bra{x}$, so the restrictions of $S_\psi$ and $S_x$ are rank-one reflections. The dynamics therefore reduce to a rotation on $\mathcal{K}$ determined by a single principal angle.

This principal angle can be constructed from the eigenvalues of $P_\psi P_x P_\psi$ restricted to the range of $P_\psi$, where $P_\psi = \ket{\psi}\bra{\psi}$ and $P_x = \ket{x}\bra{x}$ are the projectors onto the respective $-1$ eigenspaces within $\mathcal{K}$:
\begin{equation}
    P_\psi P_x P_\psi \ket{\psi} = |\braket{x|\psi}|^2 \ket{\psi}
\end{equation}
so $\cos^2(\gamma) = |\braket{x|\psi}|^2$. Defining $\sin(\theta) = \sqrt{p} = |\braket{x|\psi}|$ yields $\gamma = \pi/2 - \theta$. Creating an orthonormal basis from $\ket{\psi}$ for the rotation plane gives:
\begin{equation}
    \ket{u} = \ket{\psi}, \qquad \ket{w} = \frac{\ket{x} - \sin(\theta)\ket{\psi}}{\cos(\theta)} = \ket{\psi^\perp}
\end{equation}
in which the normalised marked component decomposes as $\ket{x} = \sin(\theta)\ket{\psi} + \cos(\theta)\ket{\psi^\perp}$. The product $Q = S_\psi S_x$ rotates states on this plane by twice the principal angle. Reflecting first about $\ket{x}$ and then about $\ket{\psi}$, the rotation angle is $2\theta - \pi$. Starting from $\ket{\psi}$, which lies entirely along $\ket{u}$ on this plane, after $t$ iterations:
\begin{equation}
    Q^t\ket{\psi} = (-1)^t\bigl(\cos(2t\theta)\ket{\psi} + \sin(2t\theta)\ket{\psi^\perp}\bigr)
    \label{eq:amp-it}
\end{equation}
up to a global phase. Since $Q^t\ket{\psi}$ remains in $\mathcal{K}$, the probability of obtaining a marked outcome equals its squared overlap with $\ket{x} = \sin(\theta)\ket{\psi} + \cos(\theta)\ket{\psi^\perp}$:
\begin{equation}
\begin{split}
    \|P_G Q^t\ket{\psi}\|^2 = |\braket{x|Q^t|\psi}|^2 &= \bigl|\sin(\theta)\cos(2t\theta) + \cos(\theta)\sin(2t\theta)\bigr|^2 = \sin^2((2t+1)\theta)
\end{split}
\end{equation}
and the iteration count nearest its first maximum is
\begin{equation}
    t^* = \left\lfloor \frac{\pi}{4\theta} - \frac{1}{2} \right\rceil = \left\lfloor \frac{\pi}{4\arcsin(\sqrt{p})} - \frac{1}{2} \right\rceil
\end{equation}
where $\lfloor\cdot\rceil$ denotes the nearest integer. When the amplified state is measured, a target outcome can therefore be obtained with constant success probability using $O(1/\sqrt{p})$ amplification iterations. Achieving the same success probability by independently preparing and measuring $\ket{\psi}$ requires $O(1/p)$ attempts, giving a quadratic improvement.

Grover's algorithm for unstructured search~\cite{Grover} is the most well-known instance of this approach. For Grover search, the state preparation algorithm is $A = H^{\otimes n}$, producing the uniform superposition $\ket{\psi} = \ket{+} = H^{\otimes n}\ket{0}$ over all $N = 2^n$ computational basis states. For a single target basis state $\ket{x}$, the overlap is $\braket{x|+} = 2^{-n/2}$, giving $\sqrt{p} = 2^{-n/2} = \sin(\theta)$ and an optimal iteration count of $t^* = O(\sqrt{N})$, a quadratic speedup over the $O(N)$ evaluations required classically.

\subsection{Partial Diffusion Operators}

The amplification iterate $Q = S_\psi S_x$ consists of two reflections. While the oracle $S_x$ is problem-specific, the diffusion operator $S_\psi = A S_0 A^\dagger$ requires an $n$-qubit controlled operation regardless of the state preparation algorithm $A$, requiring $O(n)$~\cite{mcx1, mcx2} to $O(n^2)$~\cite{mcx3} two-qubit gates depending on the available ancillae, with further overhead on hardware with limited qubit connectivity. Although the implementation cost of the diffusion operator is generally less significant than that of the oracle~\cite{quantumSearchPractical}, it is problem-independent and appears in every iteration, making it a natural target for optimisation.

Consider a partition of the $n$-qubit space into $m$ registers of sizes $n_1, \ldots, n_m$ with $\sum_{i=1}^m n_i = n$, and suppose the initial state decomposes as $\ket{\psi} = \bigotimes_{i=1}^m \ket{\psi_i}$. The partial diffusion operator on register $i$ is:
\begin{equation}
    S_{\psi_i} = \mathbb{I}_{\bar{i}} \otimes A_i\bigl(\mathbb{I}_{n_i} - 2\ket{0_i}\bra{0_i}\bigr)A_i^\dagger \otimes \mathbb{I}_{\underline{i}} = \mathbb{I}_{\bar{i}} \otimes \bigl(\mathbb{I}_{n_i} - 2\ket{\psi_i}\bra{\psi_i}\bigr) \otimes \mathbb{I}_{\underline{i}}
\end{equation}
where $\mathbb{I}_{\bar{i}}, \mathbb{I}_{\underline{i}}$ refer to each register before and after $i$. This condition is naturally satisfied by unstructured search where $\ket{\psi_i} = \ket{+_i}$, and more generally by any amplitude amplification problem where $A = \bigotimes A_i$. Each $S_{\psi_i}$ is a reflection whose reflected subspace is $\ket{\psi_i}$ tensored with the identity on all other registers, requiring only an $n_i$-qubit controlled operation rather than the full $n$-qubit one. Previous work has shown that utilising partial diffusions can significantly reduce the number of global reflections required for quantum search~\cite{grover_trade-offs_2002, searchDepthOptimization, hardware, groverOptGates, hardwareGrover2}, but at the cost of considerable analytical complexity: partial diffusions break the two-dimensional rotational structure that makes standard amplitude amplification tractable~\cite{so3, hardware}. Among these, the method of~\cite{hardware} removes the global reflection entirely by alternating two partial diffusion operators, although its query count is established only numerically. In this work we provide a recursive construction that replaces the global reflection with partial diffusion operators on any number of registers. When the projector onto the marked subspace factorises across the same registers as the state preparation unitary $A$, we derive its success probability and query count analytically, establishing when the quadratic speedup is retained.

\section{Quantum Search without Global Diffusion}
\label{sec:algorithm}
In this section we present the theoretical contributions of the work: a recursive construction for quantum search, exact expressions for its dynamics, a bound on the oracle complexity, and an application of the approach to unstructured search. Our central result is that quantum search retains its quadratic speedup with only a small subset of qubits acted on between any two successive oracle calls. We provide a recursive construction that achieves this when the state preparation unitary $A$ and the projector $P_G$ onto the marked subspace factorise across the same partition of the search qubits, with each non-oracle operator acting on a single register of this partition. Under these assumptions, we derive an exact analytical expression for the dynamics, revealing a degeneracy in the principal angles between the partial diffusion operators and the recursively constructed reflections.

Specifically, partition the $n$-qubit space into $m$ registers of sizes $n_1, n_2, \ldots, n_m$ with $\sum_{i=1}^m n_i = n$, and suppose:
\begin{equation}
    A = \bigotimes_{i=1}^m A_i, \qquad
    P_G = \bigotimes_{i=1}^m P_{G_i}
    \label{eq:decomp}
\end{equation}
where $A_i$ acts on register $i$ and $P_{G_i}$ is the orthogonal projector onto a local marked subspace. The oracle operator is the reflection $S_x = \mathbb{I}_n - 2P_G$, as defined in Subsection~\ref{subsec:qaa}. These assumptions include single product-state targets, such as computational-basis states, and multi-target searches whose target set is the Cartesian product of the local target sets associated with each register.

Let $\ket{\psi_i} = A_i\ket{0_i}$ denote the local initial state on register $i$, and define the overlap angle by $\sin(\theta_i) = \|P_{G_i}\ket{\psi_i}\|$, with $\theta_i \in (0,\pi/2)$. Then $p_i=\sin^2(\theta_i)$ is the probability that measuring register $i$ yields a state from the local marked subspace, and the initial global success probability is $p=\prod_{i=1}^{m}p_i = \prod_{i=1}^{m}\sin^2(\theta_i)$. The normalised projection of $\ket{\psi_i}$ onto this subspace is $\ket{x_i} = P_{G_i}\ket{\psi_i}/\sin(\theta_i)$, giving $\braket{x_i|\psi_i} = \sin(\theta_i)$. The factorisation then gives $\ket{x} = \bigotimes_{i=1}^m \ket{x_i}$, consistently with the definition of $\ket{x}$ in Subsection~\ref{subsec:qaa}.

For each register $i$, let $\mathcal{K}_i = \mathrm{span}\{\ket{\psi_i},\ket{x_i}\}$ denote the two-dimensional plane spanned by the local initial state and its normalised projection into the local marked subspace. Within this plane, the projector $P_{G_i}$ acts as $\ket{x_i}\bra{x_i}$, since $P_{G_i}\ket{\psi_i} = \sin(\theta_i)\ket{x_i}$ and $P_{G_i}\ket{x_i} = \ket{x_i}$. Consequently, within the tensor-product space $\mathcal{S}_m = \bigotimes_{i=1}^m \mathcal{K}_i$, the full projector $P_G$ acts as $\ket{x}\bra{x}$, and the oracle $S_x$ acts as a rank-one reflection. Both the oracle and the partial diffusion operators defined below preserve $\mathcal{S}_m$, which contains the initial state. We can therefore analyse the dynamics within this invariant subspace using the effective target state $\ket{x}$.

The partial diffusion operator on register $i$ is the conjugation of the zero-state reflection by $A_i$:
\begin{equation}
    S_{\psi_i} = A_i(\mathbb{I}_{n_i} - 2\ket{0_i}\bra{0_i})A_i^\dagger
        = \mathbb{I}_{n_i} - 2\ket{\psi_i}\bra{\psi_i}
        \label{eq:partial_diffusion}
\end{equation}
extended to the full space as a tensor product with the identity on all other registers. Each $S_{\psi_i}$ is implementable using only $A_i$, $A_i^\dagger$, and a local phase flip to the $\ket{0}$ state on register $i$. We define recursively the reflection:
\begin{equation}
    W_i = (S_{\psi_i} W_{i-1})^{t_i}\, S_{\psi_i}\, (W_{i-1} S_{\psi_i})^{t_i},
    \quad W_0 = S_x
\end{equation}
\begin{figure}[t]
\centering
\begin{quantikz}
\lstick[2]{$\ket{\psi_{m}}$}&\gate[8]{W_{i}}&\midstick[8]{\quad   =  \quad }&\gategroup[8,steps=3,style={dashed,rounded
corners,fill=blue!20, inner
xsep=2pt},background,label style={label
position=below,anchor=north,yshift=-0.2cm}]{{Repeated $t_i$ times}}&\gate[8]{W_{i-1}}&&&\gategroup[8,steps=3,style={dashed,rounded
corners,fill=blue!20, inner
xsep=2pt},background,label style={label
position=below,anchor=north,yshift=-0.2cm}]{{Repeated $t_i$ times}}&\gate[8]{W_{i-1}}&&\\
&&&&&&&&&&\\
\vdots\\
\lstick[2]{$\ket{\psi_{i}}$}&&&\gate[2]{S_{\psi_i}}&&&\gate[2]{S_{\psi_i}}&&&\gate[2]{S_{\psi_i}}&\\
&&&&&&&&&& \\
\vdots\\
\lstick[2]{$\ket{\psi_{1}}$}&&&&&&&&&&\\
&&&&&&&&&&
\end{quantikz}
\caption{\label{fig:w_decomp} Quantum circuit decomposition for the operator $W_i$. The reflection consists of $t_i$ iterates of the operators $(S_{\psi_i}W_{i-1}$), the reflection operator $S_{\psi_i}$ and then $t_i$ iterates of $(W_{i-1}S_{\psi_i})$}

\end{figure}
where $t_i \geq 1$ is the iteration count at stage $i$. The circuit structure of $W_i$ is illustrated in Figure~\ref{fig:w_decomp}. Each $W_i$ is a conjugation of $S_{\psi_i}$ by $(S_{\psi_i} W_{i-1})^{t_i}$ and is therefore itself a reflection. The product $S_{\psi_i} W_{i-1}$ is a product of two reflections whose action is characterised by the principal angles between their reflected subspaces. At stage $i$, we consider their restrictions to $\mathcal{S}_i = \bigotimes_{j=1}^i \mathcal{K}_j$, with the higher registers fixed at $\ket{x_m,\ldots,x_{i+1}}$. We prove in Subsection~\ref{sec:proof} that, despite these subspaces being high-dimensional (dimension $2^{i-1}$), the principal angles take at most two distinct values, $|\gamma_i|$ and $\pi/2 - |\gamma_i|$. Here $\gamma_i$ is a signed angle parameter defined recursively by
\begin{equation}
    \sin(2\gamma_i)
    = \sin(2\theta_i)\sin(2t_{i-1}\gamma_{i-1}),
    \quad \gamma_1 = \theta_1
\end{equation}
For $i \geq 2$, we choose $\gamma_i \in [-\pi/4,\pi/4]$ so that $|\gamma_i|$ represents the smaller of the two principal angles. As such, the squared projections onto $\ket{\psi_i}$ and its orthogonal complement $\ket{\psi_i^\perp}$ (defined within $\mathrm{span}\{\ket{\psi_i}, \ket{x_i}\}$) depend only on $\gamma_i$ and $t_i$, regardless of the state of the lower registers. Specifically, within the space where registers $m$ through $i+1$ carry their target components $\ket{x_m, \ldots, x_{i+1}}$, for any normalised $\ket{\phi_{i-1}}$ in the invariant subspace $\mathcal{S}_{i-1} = \mathrm{span}\{\ket{s_{i-1}, \ldots, s_1} : s_j \in \{\psi_j, \psi_j^\perp\}\}$:
\begin{equation}
    \|P_{\psi_i}(S_{\psi_i} W_{i-1})^{t_i}\ket{\psi_i, \phi_{i-1}}\|^2 = \cos^2(2t_i\gamma_i), \qquad
    \|P_{\psi_i^\perp}(S_{\psi_i} W_{i-1})^{t_i}\ket{\psi_i, \phi_{i-1}}\|^2 = \sin^2(2t_i\gamma_i)
    \label{eq:proj-prop}
\end{equation}
where $P_{\psi_i} = \ket{\psi_i}\bra{\psi_i} \otimes \mathbb{I}_{\mathcal{S}_{i-1}}$ and $P_{\psi_i^\perp} = \ket{\psi_i^\perp}\bra{\psi_i^\perp} \otimes \mathbb{I}_{\mathcal{S}_{i-1}}$. Applying the iterate $(S_{\psi_m} W_{m-1})^{t_m}$ to the initial state $\ket{\psi} = A\ket{0} = \bigotimes_{i=1}^m \ket{\psi_i}$ yields a state whose amplitude on register $m$ is concentrated towards $\ket{x_m}$. Since $P_{G_m}$ acts as $\ket{x_m}\bra{x_m}$ within $\mathcal{K}_m$, the squared projection onto $\ket{x_m}$ gives the success probability on register $m$. The projections in Equation~\ref{eq:proj-prop} enable a closed-form expression for this probability, which we prove in Subsection~\ref{sec:succ} is given by
\begin{equation}
    \|P_{x_m}(S_{\psi_m} W_{m-1})^{t_m}\ket{\psi}\|^2 = \frac{1}{2}\left[1
    - \frac{\cos(2\theta_m)}{\cos(2\gamma_m)}
    \cos\bigl(2(2t_m+1)\gamma_m\bigr)\right], \quad
    t_m^* = \left\lfloor\frac{\pi}{4|\gamma_m|}
    - \frac{1}{2}\right\rceil
    \label{eq:stage_succ}
\end{equation}
where $P_{x_m} = \ket{x_m}\bra{x_m}\otimes \mathbb{I}$. $t_m^*$ represents the integer nearest the first success-probability maximum when $0 < \theta_m < \pi/4$ and $\gamma_m \neq 0$. This probability is at most $\frac{1}{2}[1+\cos(2\theta_m)/\cos(2\gamma_m)]$, which is strictly less than one for $|\gamma_m|<\theta_m$. This is because $W_{m-1}$ generally entangles register $m$ with the lower registers. When $|\gamma_m|=\theta_m$, it instead acts on register $m$ as an isolated reflection about $\ket{x_m}$, and the final stage is exactly amplitude amplification on that register.
\begin{figure}[t]
\centering
\begin{quantikz}
\lstick[2]{$\ket{\psi_{m}}$}&\gategroup[8,steps=3,style={dashed,rounded
corners,fill=blue!20, inner
xsep=2pt},background,label style={label
position=below,anchor=north,yshift=-0.2cm}]{{Repeated $t_m$ times}}&\gate[8]{W_{{m-1}}}&\gate[2]{S_{\psi_m}}&\gategroup[8,steps=3,style={dashed,rounded
corners,fill=blue!20, inner
xsep=2pt},background,label style={label
position=below,anchor=north,yshift=-0.2cm}]{{For $i = m-1, \dots, 1$}}&&\gate[8]{W_{{i-1}}}&\meter{}\\
&&&&&&&\meter{}\\
\vdots\\
\lstick[2]{$\ket{\psi_{i}}$}&&&&&\gate[2]{S_{\psi_i}}&&\meter{}\\
&&&&&&&\meter{}\\
\vdots\\
\lstick[2]{$\ket{\psi_{1}}$}&&&&&&&\meter{}\\
&&&&&&&\meter{}
\end{quantikz}
\caption{ \label{fig:full_algo}Quantum circuit decomposition for the full amplification procedure. The system is initialised to the state $\ket{\psi_i}$ on each of the $m$ registers. The iterate $(S_{\psi_m}W_{m-1})$ is applied $t_m$ times, amplifying the probability of measuring $\ket{x_m}$ in the final register. A single inverse iterate $(S_{\psi_i}W_{i-1})^{-1}$ is then applied for each inner register $i = m-1, \dots, 1$, with $t_i = 1$, and all registers are measured once at the end. }

\end{figure}

After amplifying register $m$, the remaining registers are processed to recover the full target state. With $t_i=1$ at all intermediate stages, we show that the remaining target components can be recovered by applying a single inverse iterate $(S_{\psi_k}W_{k-1})^{-1}=W_{k-1}S_{\psi_k}$ at each remaining stage $k=m-1,\ldots,1$, as illustrated in Figure~\ref{fig:full_algo}. Appendix~\ref{app:run} derives the resulting state and exact success probability, from which we establish a bound on the overall success probability in Subsection~\ref{sec:overall-succ}. With these intermediate iteration counts, the oracle cost of the final-stage amplification relative to standard amplitude amplification is bounded by a factor of $1/\prod_{i=1}^{m}\cos(\theta_i)$. The inverse iterates require at most $(2^{m-1}-1)/3$ additional oracle calls, which is less than $\sqrt{N}/3$ when each partition contains at least two qubits. Applied to unstructured search of a single target state (Subsection~\ref{subsec:unstructured}), these bounds show that the approach retains the $O(\sqrt{N})$ oracle complexity of Grover search when each partition contains at least $\log_2(\log_2(N))$ qubits.

\subsection{Operator Analysis}
\label{sec:proof}
In this subsection we prove the claim in Equation~\ref{eq:proj-prop} by induction. We first outline the preliminaries before providing the full proof. For the purpose of the proof we assume that the state preparation circuit $A$ and the marked-subspace projector $P_G$ both decompose according to Equation~\ref{eq:decomp}. As established in the overview, $P_{G_i}$ acts as $\ket{x_i}\bra{x_i}$ within the plane $\mathcal{K}_i = \mathrm{span}\{\ket{\psi_i},\ket{x_i}\}$. The oracle $S_x = \mathbb{I}_n - 2P_G$ and each partial diffusion operator preserve the global space $\mathcal{S}_m = \bigotimes_{i=1}^m\mathcal{K}_i$, which contains the initial state. Within this space, $P_G$ acts as $\ket{x}\bra{x}$, where $\ket{x}=\bigotimes_{i=1}^m\ket{x_i}$. All operators and eigenspaces described below are taken to be restricted to this invariant space, with any additional restriction on the higher registers stated explicitly. In particular, for $\ket{v}\in\mathcal{S}_m$, $\|P_G\ket{v}\|^2=|\braket{x|v}|^2$.

\subsubsection{Preliminaries}
Let $S_{\psi_i}$ denote the partial diffusion operator in Equation~\ref{eq:partial_diffusion}, reflecting about the initial state $\ket{\psi_i}$ within register $i$, and let $S_x$ denote the oracle, whose restriction to $\mathcal{S}_m$ applies a phase of $-1$ to the target state $\ket{x}$:
\begin{equation}
    \left.S_x\right|_{\mathcal{S}_m} = \mathbb{I}_{\mathcal{S}_m} - 2\ket{x}\bra{x} = \mathbb{I}_{\mathcal{S}_m}
    - 2\left(\ket{x_m}\bra{x_m}\otimes \cdots
    \otimes \ket{x_1}\bra{x_1}\right)
\end{equation} 
Let $\ket{\psi_i^\perp}$ denote the orthogonal complement of $\ket{\psi_i}$ within $\mathrm{span}\{\ket{\psi_i}, \ket{x_i}\}$:
\begin{equation}
    \ket{\psi_i^\perp}
    = \frac{\ket{x_i} - \sin(\theta_i)\ket{\psi_i}}{\cos(\theta_i)},
    \quad \sin(\theta_i) = |\!\braket{x_i|\psi_i}\!|
\end{equation}
We additionally define the orthogonal complement to $\ket{x_i}$ within $\mathrm{span}\{\ket{\psi_i}, \ket{\psi_i^\perp}\}$:
\begin{equation}
    \ket{x_i^\perp} = \cos(\theta_i)\ket{\psi_i}
    - \sin(\theta_i)\ket{\psi_i^\perp}
\end{equation}
and the subspace:
\begin{equation}
    \mathcal{S}_k = \mathrm{span}\!\left\{\ket{s_k, \ldots, s_1}
    : s_j \in \{\psi_j,\, \psi_j^\perp\}\right\}
\end{equation}
which has dimension $2^k$. Each $S_{\psi_i}$ preserves $\mathrm{span}\{\ket{\psi_i}, \ket{\psi_i^\perp}\}$ at register $i$, and $S_x$ preserves $\mathcal{S}_m$. Therefore, for an $m$-register partition, $\mathcal{S}_m$ is invariant under any product of these operators.

\subsubsection{Proof}
At stage $i$, we restrict the analysis to $\ket{x_m,\ldots,x_{i+1}}\otimes\mathcal{S}_i$. On the orthogonal complement of this subspace within $\mathcal{S}_m$, the oracle acts as the identity, and each lower recursive reflection $W_j$, for $j\leq i$, reduces to $S_{\psi_j}$. We define the recursive reflection $W_i$ as:
\begin{equation}
    W_i = (S_{\psi_i} W_{i-1})^{t_i} \, S_{\psi_i} \, (W_{i-1} S_{\psi_i})^{t_i},
    \quad W_0 = S_x
\end{equation}
We aim to prove by induction that within this space defined by the correct target prefix, $(S_{\psi_i} W_{i-1})^{t_i}$ satisfies the following norms when projected onto the space spanned by $\ket{\psi_i}$ and $\ket{\psi_i^\perp}$:
\begin{equation}
    \begin{split}
        \|P_{\psi_i}(S_{\psi_i}W_{i-1})^{t_i}\ket{\psi_i,\phi_{i-1}}\|^2
        = \cos^2(2t_i\gamma_i), \quad \|P_{\psi_i^\perp}(S_{\psi_i}W_{i-1})^{t_i}\ket{\psi_i,\phi_{i-1}}\|^2
        =\sin^2(2t_i\gamma_i)
    \end{split}
    \label{equation:main_claim}
\end{equation}
for any normalised $\ket{\phi_{i-1}} \in \mathcal{S}_{i-1}$, where
$P_{\psi_i} = \ket{\psi_i}\bra{\psi_i} \otimes \mathbb{I}_{\mathcal{S}_{i-1}}$
and $P_{\psi_i^\perp} = \ket{\psi_i^\perp}\bra{\psi_i^\perp} \otimes \mathbb{I}_{\mathcal{S}_{i-1}}$
are projectors on register $i$, and $\gamma_i$ is
independent of $\ket{\phi_{i-1}}$ and defined recursively by:
\begin{equation}
    \sin(2\gamma_i) = \sin(2\theta_i)\sin(2t_{i-1}\gamma_{i-1}),
    \quad \gamma_1 = \theta_1
    \label{equation:gamma_recursion}
\end{equation}
where $\sin(\theta_i) = |\braket{x_i | \psi_i}|$, with $\theta_i \in [0,\pi/2]$ and $\gamma_i \in [-\pi/4,\pi/4]$ for $i \geq 2$. We consider the case $\theta_i \in (0,\pi/2)$ and $\sin(2t_{i-1}\gamma_{i-1}) \neq 0$ for $i \geq 2$, so that $0 < |\gamma_i| \leq \pi/4$. The key conceptual contribution is that, although $S_{\psi_i}W_{i-1}$ acts on subspaces of dimension $2^{i-1}$, its projection onto $\mathrm{span}\{\ket{\psi_i}, \ket{\psi_i^\perp}\}$ within register $i$ is fully characterised by the single angle $\gamma_i$.

\paragraph{Base Case}

At stage $1$, $W_0 = S_x$ and the iterate $(S_{\psi_1}S_x)^{t_1}$ applied to $\ket{\psi_1}$ is simply a standard amplitude amplification within this subspace, yielding the following projections, which can be derived from Equation~\ref{eq:amp-it}:
\begin{equation}
    \begin{split}
        \|P_{\psi_1}(S_{\psi_1}W_{0})^{t_1}\ket{\psi_1}\|^2
        = \cos^2(2t_1\theta_1),\quad   \|P_{\psi_1^\perp}(S_{\psi_1}W_{0})^{t_1}\ket{\psi_1}\|^2 = \sin^2(2t_1\theta_1)
    \end{split}
    \label{equation:base_case_stage1}
\end{equation}
We now analyse the effect of the iterate $(S_{\psi_2}W_{1})$ on the first two stages. \\Since $W_1 = (S_{\psi_1}W_{0})^{t_1}S_{\psi_1}(W_{0}S_{\psi_1})^{t_1}$, its reflected eigenspace is $(S_{\psi_1}W_{0})^{t_1}$ applied to the $-1$ eigenspace of $S_{\psi_1}$. On the local plane $\mathcal{K}_1$, the $-1$ eigenspace of $S_{\psi_1}$ is spanned by $\ket{\psi_1}$. Since $(S_{\psi_1}W_{0})^{t_1}\ket{\psi_1} = (-1)^{t_1}[\cos(2t_1\theta_1)\ket{\psi_1} + \sin(2t_1\theta_1)\ket{\psi_1^\perp}]$ on the branch where register $2$ is in $\ket{x_2}$, the $-1$ eigenspace of $W_1$, restricted to the first two registers in the correct branch, is:
\begin{equation}
    \mathcal{E}_{W_1} = \mathrm{span}\bigl\{\ket{x_2}
    \otimes (S_{\psi_1}W_{0})^{t_1}\ket{\psi_{1}},\quad
    \ket{x_2^\perp}\otimes\ket{\psi_{1}}\bigr\}
\end{equation}
Similarly, the $-1$ eigenspace of $S_{\psi_2}$ is:
\begin{equation}
    \mathcal{E}_{S_2} = \mathrm{span}\bigl\{\ket{\psi_2} \otimes \mathcal{S}_{1} \bigr\}
\end{equation}
Let $P_S$ and $P_W$ denote the orthogonal projectors onto $\mathcal{E}_{S_2}$ and $\mathcal{E}_{W_1}$ respectively:
\begin{equation}
    P_S = \ket{\psi_2}\bra{\psi_2}
    \otimes \mathbb{I}_{\mathcal{S}_{1}}
\end{equation}
\begin{equation}
    P_W = \ket{x_2}\bra{x_2} \otimes Q_{1}
    + \ket{x_2^\perp}\bra{x_2^\perp} \otimes P_\psi
\end{equation}
where $P_\psi = \ket{\psi_{1}}\bra{\psi_{1}}$ is the projector onto $\ket{\psi_{1}}$ within $\mathcal{S}_{1}$, and:
\begin{equation}
    Q_{1} = (S_{\psi_1}W_{0})^{t_1} P_\psi (W_{0}S_{\psi_1})^{t_1}
\end{equation}
is the projector onto the image of $\ket{\psi_{1}}$ under $(S_{\psi_1}W_{0})^{t_1}$. Since $\mathcal{S}_1$ is two-dimensional, $Q_1$ is a scalar $2 \times 2$ matrix in the $\{\ket{\psi_1}, \ket{\psi_1^\perp}\}$ basis. Using (\ref{eq:amp-it}), with the global phase cancelling in the projector:
\begin{equation}
    Q_{1} = \begin{pmatrix}
        \cos^2(2t_{1}\theta_{1}) &  \cos(2t_{1}\theta_{1})\sin(2t_{1}\theta_{1})  \\
        \cos(2t_{1}\theta_{1})\sin(2t_{1}\theta_{1})
        &  \sin^2(2t_{1}\theta_{1})
    \end{pmatrix}
\end{equation}
The principal angles between the two eigenspaces are determined by the eigenvalues of $P_S P_W P_S$ restricted to $\mathcal{E}_{S_2}$. For any $\ket{\phi} \in \mathcal{S}_{1}$:
\begin{equation}
\begin{split}
    P_S P_W P_S \ket{\psi_2}\ket{\phi}
    &= P_S\Bigl[\sin(\theta_2)\ket{x_2}\, Q_{1}\ket{\phi}
    + \cos(\theta_2)\ket{x_2^\perp}\, P_\psi\ket{\phi}\Bigr] \\
    &= \ket{\psi_2}\Bigl[\sin^2(\theta_2)\, Q_{1}
    + \cos^2(\theta_2)\, P_\psi\Bigr]\ket{\phi}
\end{split}
\label{eq:operator_eigenvalues}
\end{equation}
The eigenvalues of $P_S P_W P_S|_{\mathcal{E}_{S_2}}$ are therefore those of the operator on $\mathcal{S}_{1}$:
\begin{equation}
    T = \sin^2(\theta_2)\, Q_{1} + \cos^2(\theta_2)\, P_\psi
\end{equation}
We decompose $\mathcal{S}_{1} = \mathrm{range}(P_\psi) \oplus \mathrm{range}(P_{\psi^\perp})$ where $P_{\psi^\perp} = \ket{\psi_{1}^\perp}\bra{\psi_{1}^\perp}$ and $P_\psi + P_{\psi^\perp} = \mathbb{I}_{\mathcal{S}_{1}}$. Substituting the entries of $Q_1$:
\begin{equation}
    T = \begin{pmatrix}
        \sin^2(\theta_2)\cos^2(2t_{1}\theta_{1})  + \cos^2(\theta_2) &  \sin^2(\theta_2)\cos(2t_{1}\theta_{1})\sin(2t_{1}\theta_{1})  \\
        \sin^2(\theta_2)\cos(2t_{1}\theta_{1})\sin(2t_{1}\theta_{1})
        &  \sin^2(\theta_2)\sin^2(2t_{1}\theta_{1})
    \end{pmatrix}
    \label{eq:t_2}
\end{equation}
with eigenvalues:
\begin{equation}
    \lambda_\pm = \frac{1 \pm \sqrt{1
    - \sin^2(2\theta_2)\sin^2(2t_{1}\theta_1)}}{2}
\end{equation}
Defining $\gamma_2$ by $\sin(2\gamma_2) = \sin(2\theta_2)\sin(2t_1\theta_1)$, with the same constraints as in Equation~\ref{equation:gamma_recursion}, the eigenvalues become $\lambda_+ = \cos^2(\gamma_2)$ and $\lambda_- = \sin^2(\gamma_2)$. The principal angles between $\mathcal{E}_{W_1}$ and $\mathcal{E}_{S_2}$ are therefore $|\gamma_2|$ and $\pi/2 - |\gamma_2|$, with rotation magnitudes $2|\gamma_2|$ and $\pi - 2|\gamma_2|$. The product $S_{\psi_2} W_{1} = (\mathbb{I} - 2P_S)(\mathbb{I} - 2P_W)$ decomposes into two invariant two-dimensional planes, one associated with principal angle $|\gamma_2|$ (rotation magnitude $2|\gamma_2|$) and one associated with principal angle $\pi/2 - |\gamma_2|$ (rotation magnitude $\pi - 2|\gamma_2|$). In each plane, the rotation acts between a vector $\ket{u}$ in $\mathcal{E}_{S_2}$ and a vector $\ket{w}$ orthogonal to $\mathcal{E}_{S_2}$ within that plane. By construction:
\begin{equation}
    \|P_{\psi_2} \ket{u}\|^2 = 1, \quad
    \|P_{\psi_2} \ket{w}\|^2 = 0, \quad
    \|P_{\psi_2^\perp} \ket{u}\|^2 = 0, \quad
    \|P_{\psi_2^\perp} \ket{w}\|^2 = 1
    \label{eq:projections}
\end{equation}
The eigenvalues determine $|\gamma_2|$ but do not specify an orientation for the rotation planes. We choose $\ket{w} = -2(\mathbb{I} - P_S)P_W\ket{u}/\sin(2\gamma_2)$, giving rotation angles $2\gamma_2$ and $\pi - 2\gamma_2$ in the respective planes. As we show below, the squared projections onto $\ket{\psi_2}$ and $\ket{\psi_2^\perp}$ are independent of the sign of $\gamma_2$.

Any state $\ket{\psi_2}\ket{\phi}$ with normalised $\ket{\phi} \in \mathcal{S}_{1}$ lies in $\mathcal{E}_{S_2}$ and therefore decomposes as $\ket{\psi_2}\ket{\phi} = \alpha \ket{u^{(1)}} + \beta \ket{u^{(2)}}$ where the superscripts $(1)$ and $(2)$ denote the two planes, with $|\alpha|^2 + |\beta|^2 = 1$. After $t_2$ applications of $S_{\psi_2} W_{1}$, in the plane with rotation angle $2\gamma_2$:
\begin{equation}
    (S_{\psi_2} W_{1})^{t_2}\ket{u^{(1)}}
    = \cos(2t_2\gamma_2)\ket{u^{(1)}}
    + \sin(2t_2\gamma_2)\ket{w^{(1)}}
\end{equation}
and in the plane with rotation angle $\pi - 2\gamma_2$:
\begin{equation}
    (S_{\psi_2} W_{1})^{t_2}\ket{u^{(2)}}
    = (-1)^{t_2}\bigl[\cos(2t_2\gamma_2)\ket{u^{(2)}}
    - \sin(2t_2\gamma_2)\ket{w^{(2)}}\bigr]
\end{equation}
Projecting onto $\ket{\psi_2}$ and using orthogonality of $\ket{u^{(1)}}$ and $\ket{u^{(2)}}$ across different planes together with $\|P_{\psi_2}\ket{u}\|^2 = 1$:
\begin{equation}
    \|P_{\psi_2}(S_{\psi_2} W_{1})^{t_2}\ket{\psi_2, \phi}\|^2
    = \cos^2(2t_2\gamma_2)
\end{equation}
Similarly, $\|P_{\psi_2^\perp}(S_{\psi_2} W_{1})^{t_2}\ket{\psi_2, \phi}\|^2 = \sin^2(2t_2\gamma_2)$, establishing the claim at stage $2$ with $\gamma_2$ independent of $\ket{\phi}$ and satisfying $\sin(2\gamma_2) = \sin(2\theta_2)\sin(2t_1\theta_1)$, confirming the recursion~\eqref{equation:gamma_recursion} with $\gamma_1 = \theta_1$.
\paragraph{Inductive Step} We assume the claim holds at stage $i{-}1$: for any normalised $\ket{\phi_{i-2}} \in \mathcal{S}_{i-2}$ and on the subspace where registers $m$ through $i$ carry their target values,
\begin{equation}
    \begin{split}
        \|P_{\psi_{i-1}}(S_{\psi_{i-1}}W_{i-2})^{t_{i-1}}
        \ket{\psi_{i-1},\phi_{i-2}}\|^2
        &= \cos^2(2t_{i-1}\gamma_{i-1})\\
        \|P_{\psi_{i-1}^\perp}(S_{\psi_{i-1}}W_{i-2})^{t_{i-1}}
        \ket{\psi_{i-1},\phi_{i-2}}\|^2
        &= \sin^2(2t_{i-1}\gamma_{i-1})
    \end{split}
    \label{equation:inductive_hypothesis}
\end{equation}
The analysis of the iterate $(S_{\psi_i}W_{i-1})$ on registers $i$ through $1$ follows the same structure as the base case. Since $W_{i-1} = (S_{\psi_{i-1}}W_{i-2})^{t_{i-1}} S_{\psi_{i-1}} (W_{i-2}S_{\psi_{i-1}})^{t_{i-1}}$, its $-1$ eigenspace is $(S_{\psi_{i-1}}W_{i-2})^{t_{i-1}}$ applied to the $-1$ eigenspace of $S_{\psi_{i-1}}$. Unlike the base case, the $-1$ eigenspace of $S_{\psi_{i-1}}$ is $\ket{\psi_{i-1}} \otimes \mathcal{S}_{i-2}$, which has dimension $2^{i-2}$ rather than $1$. The $-1$ eigenspaces of $W_{i-1}$ and $S_{\psi_i}$, restricted to registers $i$ through $1$ in the correct branch, are:
\begin{equation}
    \mathcal{E}_{W} = \mathrm{span}\bigl\{\ket{x_i}
    \otimes (S_{\psi_{i-1}}W_{i-2})^{t_{i-1}}(\ket{\psi_{i-1}} \otimes \mathcal{S}_{i-2}),\;
    \ket{x_i^\perp}\otimes\ket{\psi_{i-1}}
    \otimes \mathcal{S}_{i-2}\bigr\}, \quad \mathcal{E}_{S} = \ket{\psi_i} \otimes \mathcal{S}_{i-1}
    \label{eq:spaces}
\end{equation}
The corresponding projectors onto these eigenspaces are:
\begin{equation}
    P_S = \ket{\psi_i}\bra{\psi_i}
    \otimes \mathbb{I}_{\mathcal{S}_{i-1}}, \qquad
    P_W = \ket{x_i}\bra{x_i} \otimes Q_{i-1}
    + \ket{x_i^\perp}\bra{x_i^\perp} \otimes P_\psi
\end{equation}
where $P_\psi = \ket{\psi_{i-1}}\bra{\psi_{i-1}} \otimes \mathbb{I}_{\mathcal{S}_{i-2}}$ is the projector onto $\ket{\psi_{i-1}} \otimes \mathcal{S}_{i-2}$ within $\mathcal{S}_{i-1}$, and:
\begin{equation}
    Q_{i-1} = (S_{\psi_{i-1}}W_{i-2})^{t_{i-1}} \, P_\psi \, (W_{i-2}S_{\psi_{i-1}})^{t_{i-1}}
\end{equation}
is the projector onto the image of $\ket{\psi_{i-1}} \otimes \mathcal{S}_{i-2}$ under $(S_{\psi_{i-1}}W_{i-2})^{t_{i-1}}$. Following from the base case in Equation~\ref{eq:operator_eigenvalues}, the eigenvalues of $P_S P_W P_S|_{\mathcal{E}_S}$ are those of the operator on $\mathcal{S}_{i-1}$:
\begin{equation}
    T = \sin^2(\theta_i)\, Q_{i-1} + \cos^2(\theta_i)\, P_\psi
    \label{eq:T}
\end{equation}
Unlike the base case, where $Q_1$ was a scalar $2 \times 2$ matrix, $Q_{i-1}$ is now an operator on the $2^{i-1}$-dimensional space $\mathcal{S}_{i-1}$. We show that its block structure nevertheless reduces to a scalar $2 \times 2$ problem. We decompose $\mathcal{S}_{i-1} = \mathrm{range}(P_\psi) \oplus \mathrm{range}(P_{\psi^\perp})$ where $P_{\psi^\perp} = \ket{\psi_{i-1}^\perp}\bra{\psi_{i-1}^\perp} \otimes \mathbb{I}_{\mathcal{S}_{i-2}}$ and $P_\psi + P_{\psi^\perp} = \mathbb{I}_{\mathcal{S}_{i-1}}$. In this decomposition, $Q_{i-1}$ takes the block form:
\begin{equation}
    Q_{i-1} = \begin{pmatrix}
        P_\psi Q_{i-1} P_\psi & C \\
        C^\dagger & P_{\psi^\perp} Q_{i-1} P_{\psi^\perp}
    \end{pmatrix}
\end{equation}
where $C = P_\psi Q_{i-1} P_{\psi^\perp}$. We derive each block using the inductive hypothesis.

\subparagraph{Diagonal blocks.} Any normalised $\ket{\phi} \in \mathrm{range}(P_\psi)$ has the form $\ket{\psi_{i-1}}\ket{\phi_{i-2}}$ for some $\ket{\phi_{i-2}} \in \mathcal{S}_{i-2}$. By the inductive hypothesis~\eqref{equation:inductive_hypothesis}:
\begin{equation}
    \|P_\psi (S_{\psi_{i-1}}W_{i-2})^{t_{i-1}} \ket{\phi}\|^2
    = \cos^2(2t_{i-1}\gamma_{i-1}),
    \qquad
    \|P_{\psi^\perp} (S_{\psi_{i-1}}W_{i-2})^{t_{i-1}} \ket{\phi}\|^2
    = \sin^2(2t_{i-1}\gamma_{i-1})
\end{equation}
Crucially, these norms are independent of the choice of $\ket{\phi_{i-2}}$. \\The restricted map $P_\psi (S_{\psi_{i-1}}W_{i-2})^{t_{i-1}}\big|_{\mathrm{range}(P_\psi)} : \mathrm{range}(P_\psi) \to \mathrm{range}(P_\psi)$ therefore has constant output norm $|\cos(2t_{i-1}\gamma_{i-1})|$ over all normalised inputs, making it a scaled isometry. Since the domain and codomain have equal dimension $2^{i-2}$, it is a scaled unitary:
\begin{equation}
    P_\psi (S_{\psi_{i-1}}W_{i-2})^{t_{i-1}}\big|_{\mathrm{range}(P_\psi)} = \cos(2t_{i-1}\gamma_{i-1})\, V_1
\end{equation}
for some unitary $V_1$ on $\mathrm{range}(P_\psi)$. Therefore:
\begin{equation}
    P_\psi Q_{i-1} P_\psi
    = \cos^2(2t_{i-1}\gamma_{i-1})\, \mathbb{I}_{P_\psi}
    \label{equation:Q_diag_psi}
\end{equation}
By the same argument:
\begin{equation}
    P_{\psi^\perp} (S_{\psi_{i-1}}W_{i-2})^{t_{i-1}}\big|_{\mathrm{range}(P_\psi)} = \sin(2t_{i-1}\gamma_{i-1})\, V_2
\end{equation}
for some isometry $V_2$ mapping $\mathrm{range}(P_\psi)$ to $\mathrm{range}(P_{\psi^\perp})$. Since both spaces have dimension $2^{i-2}$, $V_2$ is a unitary, giving:
\begin{equation}
    P_{\psi^\perp} Q_{i-1} P_{\psi^\perp}
    = \sin^2(2t_{i-1}\gamma_{i-1})\, \mathbb{I}_{P_{\psi^\perp}}
    \label{equation:Q_diag_perp}
\end{equation}
This is the key step: the state-independence guaranteed by the inductive hypothesis forces the diagonal blocks of $Q_{i-1}$ to be scalar multiples of the identity, despite $Q_{i-1}$ being an operator on a $2^{i-1}$-dimensional space.

\subparagraph{Off-diagonal product.} Since $Q_{i-1}$ is a projector ($Q_{i-1}^2 = Q_{i-1}$), computing the top-left block of this identity in the $P_\psi/P_{\psi^\perp}$ decomposition:
\begin{equation}
    P_\psi Q_{i-1}^2 P_\psi
    = (P_\psi Q_{i-1} P_\psi)^2
    + CC^\dagger
    = P_\psi Q_{i-1} P_\psi
\end{equation}
Substituting~\eqref{equation:Q_diag_psi}:
\begin{equation}
    CC^\dagger
    = \cos^2(2t_{i-1}\gamma_{i-1})\sin^2(2t_{i-1}\gamma_{i-1})\,
    \mathbb{I}_{P_\psi}
    \label{equation:off_diag_product}
\end{equation}
which is a scalar multiple of the identity over $P_{\psi}$. $C^\dagger C$ similarly gives \\$\cos^2(2t_{i-1}\gamma_{i-1})\sin^2(2t_{i-1}\gamma_{i-1})\, \mathbb{I}_{P_{\psi^\perp}}$
\subparagraph{Eigenvalues of $T$.} With $P_\psi$ taking the form $\mathrm{diag}(\mathbb{I}_{P_\psi}, 0)$ in this decomposition, the operator $T = \sin^2(\theta_i)\, Q_{i-1} + \cos^2(\theta_i)\, P_\psi$ has the block form:
\begin{equation}
    T = \begin{bmatrix}
        d_+ \cdot \mathbb{I}_{P_\psi}
        & \sin^2(\theta_i)\, C \\[4pt]
        \sin^2(\theta_i)\, C^\dagger
        & d_- \cdot \mathbb{I}_{P_{\psi^\perp}}
    \end{bmatrix}
\end{equation}
where:
\begin{equation}
    d_+ = \sin^2(\theta_i)\cos^2(2t_{i-1}\gamma_{i-1})
    + \cos^2(\theta_i), \qquad
    d_- = \sin^2(\theta_i)\sin^2(2t_{i-1}\gamma_{i-1})
\end{equation}
Since the diagonal blocks are scalar multiples of the identity and $CC^\dagger = c\,\mathbb{I}_{P_\psi}$ by~\eqref{equation:off_diag_product}, we have $C = \sqrt{c}\,V$ for some unitary $V$. The sign of the scalar factor can be absorbed into $V$. The eigenvalue problem reduces to the scalar $2 \times 2$ matrix, with each eigenvalue having multiplicity $2^{i-2}$:
\begin{equation}
    \begin{bmatrix}
        d_+ & \sin^2(\theta_i)\cos(2t_{i-1}\gamma_{i-1})
        \sin(2t_{i-1}\gamma_{i-1}) \\[4pt]
        \sin^2(\theta_i)\cos(2t_{i-1}\gamma_{i-1})
        \sin(2t_{i-1}\gamma_{i-1}) & d_-
    \end{bmatrix}
    \label{eq:hermetian}
\end{equation}
This has the same form as the base case, with $\theta_1$ replaced by $\gamma_{i-1}$. The eigenvalues are:
\begin{equation}
    \lambda_\pm = \frac{1 \pm \sqrt{1
    - \sin^2(2\theta_i)\sin^2(2t_{i-1}\gamma_{i-1})}}{2}
\end{equation}
Defining $\gamma_i$ by $\sin(2\gamma_i) = \sin(2\theta_i)\sin(2t_{i-1}\gamma_{i-1})$ with the same constraints as in Equation~\ref{equation:gamma_recursion}, the eigenvalues become $\lambda_+ = \cos^2(\gamma_i)$ and $\lambda_- = \sin^2(\gamma_i)$. The principal angles between $\mathcal{E}_W$ and $\mathcal{E}_S$ are therefore $|\gamma_i|$ and $\pi/2 - |\gamma_i|$, each with multiplicity $2^{i-2}$, with rotation magnitudes $2|\gamma_i|$ and $\pi - 2|\gamma_i|$.

\subparagraph{Squared projections.} The product $S_{\psi_i} W_{i-1} = (\mathbb{I} - 2P_S)(\mathbb{I} - 2P_W)$ decomposes into $2^{i-2}$ invariant two-dimensional planes associated with principal angle $|\gamma_i|$ (rotation by $2\gamma_i$) and $2^{i-2}$ planes associated with $\pi/2 - |\gamma_i|$ (rotation by $\pi - 2\gamma_i$). As in the base case, in each plane the rotation acts between a vector $\ket{u_k}$ in $\mathcal{E}_{S}$ and the vector $\ket{w_k}$ of Equation~\ref{eq:w_k} (with $\gamma_i$ in place of $\gamma_m$), orthogonal to $\mathcal{E}_{S}$ within that plane, satisfying the same projection properties as Equation~\ref{eq:projections}.

Any state $\ket{\psi_i}\ket{\phi_{i-1}}$ with normalised $\ket{\phi_{i-1}} \in \mathcal{S}_{i-1}$ lies in $\mathcal{E}_{S}$ and decomposes as $\ket{\psi_i}\ket{\phi_{i-1}} = \sum_k (\alpha_k \ket{u_k^{(1)}} + \beta_k \ket{u_k^{(2)}})$ where the superscripts $(1)$ and $(2)$ denote the two families of planes, with $\sum_k(|\alpha_k|^2 + |\beta_k|^2) = 1$. Projecting onto $\ket{\psi_i}$ after $t_i$ iterations:
\begin{equation}
    P_{\psi_i}(S_{\psi_i} W_{i-1})^{t_i}\ket{\psi_i,\phi_{i-1}}
    = \cos(2t_i\gamma_i)\sum_k\left[\alpha_k\ket{u_k^{(1)}}+(-1)^{t_i}\beta_k\ket{u_k^{(2)}}\right]
\end{equation}
Taking the squared norm, and using orthogonality of the $\ket{u_k}$ across different invariant planes together with $\|P_{\psi_i}\ket{u_k}\|^2 = 1$:
\begin{equation}
    \|P_{\psi_i}(S_{\psi_i} W_{i-1})^{t_i}\ket{\psi_i,\phi_{i-1}}\|^2
    = \cos^2(2t_i\gamma_i)
\end{equation}
Similarly, $\|P_{\psi_i^\perp}(S_{\psi_i} W_{i-1})^{t_i}\ket{\psi_i,\phi_{i-1}}\|^2 = \sin^2(2t_i\gamma_i)$, establishing the claim at stage $i$ with $\gamma_i$ independent of $\ket{\phi_{i-1}}$, completing the inductive step.
\subsection{Success Probability at the Final Register}
\label{sec:succ}

Having established the projections in Equation~\ref{eq:proj-prop}, we can derive a closed form solution to the probability of measuring the target within the final register $m$ after applying $(S_{\psi_m}W_{m-1})^{t_m}$ to $\ket{\psi_m}\ket{\phi_{m-1}}$ for any normalised $\ket{\phi_{m-1}} \in \mathcal{S}_{m-1}$. We note that in this final iterate, the relevant invariant space is $\mathcal{S}_m$, and there are no higher registers to fix in their marked components. The natural choice is $\ket{\phi_{m-1}} = \ket{\psi_{m-1}, \ldots, \psi_1}$, giving the full initial state $\ket{\psi} = A\ket{0} = \bigotimes_{i=1}^m \ket{\psi_i}$, which requires no prior knowledge of $x$. Let $\ket{v} = (S_{\psi_m} W_{m-1})^{t_m}\ket{\psi}$, then the probability of obtaining a local target outcome is $\|(P_{G_m}\otimes\mathbb{I})\ket{v}\|^2=\|P_{x_m}\ket{v}\|^2$.  Expanding $\bra{x_m} = \sin(\theta_m)\bra{\psi_m} + \cos(\theta_m)\bra{\psi_m^\perp}$ and noting that $(\bra{\psi_m}\otimes\mathbb{I})\ket{v}$ and $(\bra{\psi_m^\perp}\otimes\mathbb{I})\ket{v}$ are residual vectors in $\mathcal{H}_{m-1,\ldots,1}$ with norms $|\cos(2t_m\gamma_m)|$ and $|\sin(2t_m\gamma_m)|$, respectively:
\begin{equation}
\begin{split}
    \|P_{x_m} \ket{v}\|^2
    &= \frac{1}{2}\bigl(1
    - \cos(2\theta_m)\cos(4t_m\gamma_m)\bigr) + \sin(2\theta_m)\,\mathrm{Re}\!\left[
    \bra{v}(\ket{\psi_m}\bra{\psi_m^\perp}\otimes\mathbb{I})\ket{v}\right]
\end{split}
\end{equation}
The terms outside the real part are determined by the squared projection norms onto $\ket{\psi_m}$ and $\ket{\psi_m^\perp}$. For the following derivation assume $m\geq2$ and $0<|\gamma_m|<\pi/4$, so the denominators are nonzero. To evaluate the cross term, we use the eigenvectors $\ket{\phi_k}$ of $T$ from Equation~\ref{eq:T} and the corresponding states $\ket{u_k}=\ket{\psi_m}\ket{\phi_k}$ (with eigenvalue $\cos^2(\gamma_m)$; the same formula applies to the $\sin^2(\gamma_m)$ family) and the orthogonal vectors $\ket{w_k}$ in each rotation plane:
\begin{equation}
    \ket{w_k} = \frac{-2}{\sin(2\gamma_m)}
    (\mathbb{I} - P_S)P_W\ket{u_k}
    \label{eq:w_k}
\end{equation}
Projecting onto $\ket{\psi_m^\perp}$ and using $P_{W} = \ket{x_m}\bra{x_m} \otimes Q_{m-1}
+ \ket{x_m^\perp}\bra{x_m^\perp} \otimes P_\psi$:
\begin{equation}
    \braket{\psi_m^\perp|w_k}
    = \frac{-\sin(2\theta_m)}{\sin(2\gamma_m)}
    (Q_{m-1} - P_\psi)\ket{\phi_k}
    \label{eq:w_lower}
\end{equation}
$\mathrm{Re}\!\left[\bra{v}(\ket{\psi_m}\bra{\psi_m^\perp}\otimes\mathbb{I})\ket{v}\right]$ for an initial state $\ket{u_k}$ in the first set of eigenvectors is $\sin(2t_m\gamma_m)$ $\cos(2t_m\gamma_m)\,\mathrm{Re}\braket{\phi_k|w_k^{(\mathrm{lower})}}$. It therefore remains to evaluate $\braket{\phi_k|(Q_{m-1} - P_\psi)|\phi_k}$. Explicitly diagonalising $T$ in the $P_\psi / P_{\psi^\perp}$ decomposition, the eigenvector with eigenvalue $\cos^2(\gamma_m)$ takes the form:
\begin{equation}
    \ket{\phi_k} = \cos(\xi/2)\ket{\psi_{m-1}}\ket{a_k} + \sin(\xi/2)\ket{\psi_{m-1}^\perp}\ket{b_k}
\end{equation}
where $\xi$ is obtained by diagonalising the resulting $2\times 2$ Hermitian matrix of $T$ in Equation~\ref{eq:hermetian}. Using the block structure of $Q_{m-1}$ with $\phi = 2t_{m-1}\gamma_{m-1}$ and the relations $C\ket{b_k} = \cos\phi\sin\phi\,\ket{a_k}$, $C^\dagger\ket{a_k} = \cos\phi\sin\phi\,\ket{b_k}$:
\begin{equation}
    (Q_{m-1} - P_\psi)\ket{\phi_k} = \sin\phi\bigl[
    \sin(\xi/2 - \phi)\ket{\psi_{m-1}}\ket{a_k}
    + \cos(\xi/2 - \phi)\ket{\psi_{m-1}^\perp}\ket{b_k}\bigr]
\end{equation}
Taking the inner product and applying the sine addition formula:
\begin{equation}
    \braket{\phi_k|(Q_{m-1} - P_\psi)|\phi_k}
    = \sin\phi\sin(\xi - \phi)
\end{equation}
Explicitly diagonalising Equation~\ref{eq:hermetian} yields
$\cos\xi = (1 - 2\sin^2\theta_m\sin^2\phi)/\cos(2\gamma_m)$ and\\
$\sin\xi = \sin^2\theta_m\sin(2\phi)/\cos(2\gamma_m)$, giving
\begin{equation}
    \sin(\xi - \phi)
    = \frac{-\sin\phi\cos(2\theta_m)}{\cos(2\gamma_m)}
\end{equation}
Combining:
\begin{equation}
\begin{split}
      \braket{\phi_k|w_k^{(\mathrm{lower})}}
    &= \frac{-\sin(2\theta_m)}{\sin(2\gamma_m)} \cdot \sin(\phi) \cdot\frac{-\sin\phi\cos(2\theta_m)}{\cos(2\gamma_m)} \\
    &= \frac{\sin\phi\cos(2\theta_m)}{\cos(2\gamma_m)}\\
    &= \frac{\sin(2\gamma_m)\cos(2\theta_m)}{\sin(2\theta_m)\cos(2\gamma_m)}
\end{split}
\end{equation}
The same contribution to the real cross term holds for the other family of rotation planes, since replacing $\xi/2$ with $\xi/2 + \pi/2$ flips both $\sin(\xi - \phi)$ and the relative sine coefficient. Mixed terms between the two families have zero real contribution, as $Q_{m-1}-P_\psi$ in Equation~\ref{eq:w_lower} is Hermitian and the relative sine coefficients have opposite signs, so this also holds for any normalised superposition. The cross term is therefore:
\begin{equation}
    \mathrm{Re}\!\left[\bra{v}(\ket{\psi_m}\bra{\psi_m^\perp}\otimes\mathbb{I})\ket{v}
    \right]
    = \frac{\sin(4t_m\gamma_m)\sin(2\gamma_m)\cos(2\theta_m)}{2\sin(2\theta_m)\cos(2\gamma_m)}
\end{equation}
Substituting and applying $\cos A\cos B - \sin A\sin B = \cos(A+B)$:
\begin{equation}
    \boxed{P(x_m) = \frac{1}{2}\left[1
    - \frac{\cos(2\theta_m)}{\cos(2\gamma_m)}
    \cos\bigl(2(2t_m+1)\gamma_m\bigr)\right], \quad
    t_m^* = \left\lfloor\frac{\pi}{4|\gamma_m|}
    - \frac{1}{2}\right\rceil}
    \label{eq:stage-prob}
\end{equation}
giving a closed-form expression for $P(x_m)$ that depends only on $\theta_m,\gamma_m$, and $t_m$. Here $P(x_m)$ denotes the probability of a local target outcome.

\subsection{Overall Success Probability}
\label{sec:overall-succ}
Having amplified the target component in register $m$, we now consider the probability of recovering the full target state. We apply $(S_{\psi_k}W_{k-1})^{-t_k}=(W_{k-1}S_{\psi_k})^{t_k}$ at each remaining stage $k=m-1,\ldots,1$, using the same iteration counts as in the recursive construction of $W_{m-1}$. Throughout this subsection, we assume $m\geq2$, $0<\theta_i<\pi/4$ at every register and $\sin(2t_i\gamma_i)\neq0$ for $i<m$.

Appendix~\ref{app:run} shows that, after the final-stage amplification and projection onto the target component $\ket{x_m}$, the unnormalised lower-register state is a linear combination of the initial state and its component outside the range of $Q_{m-1}$:
\begin{equation}
\begin{split}
    (\bra{x_m}\otimes\mathbb{I})
    (S_{\psi_m}W_{m-1})^{t_m}\ket{\psi}
    ={}&\mu_{m-1}(\mathbb{I}-Q_{m-1})
    \ket{\psi_{m-1},\ldots,\psi_1}\\
    &+\nu_{m-1}\ket{\psi_{m-1},\ldots,\psi_1}.
\end{split}
\end{equation}
Here $Q_{m-1}$ is the projector defined in Subsection~\ref{sec:proof}, and the real coefficients $\mu_{m-1}$ and $\nu_{m-1}$ are given in Equation~\ref{eq:end-coef}. Applying the inverse iterates and projecting onto each successive local target preserves this two-component form, with the coefficients updated by Equation~\ref{eq:app-rec}. Continuing to the first register gives the amplitude for the full target $\ket{x}$, whose squared magnitude is the overall success probability $P^*(x)$.

A simpler bound follows when $t_i$ is odd for $2\leq i<m$. Following the component with coefficient $\mu_k$, the inverse iterates leave register $k$ in $\ket{\psi_k^\perp}$. Projection onto $\ket{x_k}$ then contributes a factor $\cos(\theta_k)$, while the remaining registers retain the same recursive form. Repeating this process gives a product of local overlap factors, with a correction from the second component that we bound below.

For this, we first bound the probability of obtaining the target outcome on register $m$. Taking $t_m=t_m^*$, rounding to the nearest first maximum gives $\left|(2t_m^*+1)|\gamma_m|-\pi/2\right|\leq|\gamma_m|$. Equation~\ref{eq:stage-prob} therefore yields
\begin{equation}
    P(x_m) \geq \frac{1}{2}\bigl[1+\cos(2\theta_m)\bigr]
    = \cos^2(\theta_m).
\end{equation}
For this choice of $t_m$, Appendix~\ref{app:run} also bounds the residual factor $|\nu_{m-1}|\leq\sin(\theta_m)\tan(2|\gamma_m|)$. This, along with a product of local overlap factors gives
\begin{equation}
    \left|
    \sqrt{P^*(x)}
    -\sqrt{P(x_m)}\prod_{k=1}^{m-1}\cos(\theta_k)
    \right|
    \leq2\sin(\theta_m)\tan(2|\gamma_m|).
\end{equation}
Combining this with the final-register bound gives
\begin{equation}
    \sqrt{P^*(x)}\geq
    \prod_{i=1}^{m}\cos(\theta_i)
    -2\sin(\theta_m)\tan(2|\gamma_m|).
    \label{eq:overall-succ}
\end{equation}
As the subsequent inverse iterates preserve the target subspace of register $m$, the overall success probability is also bounded above by $P(x_m)\leq\frac{1}{2}[1+\cos(2\theta_m)/\cos(2\gamma_m)]$ (Equation~\ref{eq:stage-prob}). 
The probability of recovering the full target depends on the accumulated loss across the registers and the residual correction. Writing $p_i=\sin^2(\theta_i)$ for the initial local success probabilities, if $\sum_i p_i\leq\kappa$ for a fixed constant $\kappa$, then
\begin{equation}
    \prod_{i=1}^{m}\cos^2(\theta_i)
    =\prod_{i=1}^{m}(1-p_i)\geq4^{-\kappa},
\end{equation}
using $-\log(1-p_i)\leq2p_i\log2$ for $0<p_i<1/2$. Under this bound, for odd intermediate counts and $t_m=t_m^*$, the residual correction vanishes as $|\gamma_m|\to0$, so recovering the remaining registers preserves a nonzero success probability as the search problem grows. If $\sum_i p_i\to0$, both the accumulated loss and the residual correction vanish, giving $P^*(x)\to1$.

\subsection[Oracle Complexity]{Oracle Complexity}
\label{subsec:oracle}

To establish the overall oracle complexity, we first evaluate the oracle complexity of the final stage of the search algorithm --- that is, the complexity of amplifying register $m$ towards $\ket{x_m}$. Each reflection $W_i$ makes $2t_i$ calls to $W_{i-1}$. As $W_0 = S_x$, each operator $W_i$ has a query complexity of $C(W_i) = 2^i\prod_{j=1}^{i}t_j$ and the total query complexity of $t_m$ iterations of $W_{m-1}$ at the final stage is $C\bigl((S_{\psi_m}W_{m-1})^{t_m}\bigr) = t_m\,C(W_{m-1}) = 2^{m-1}\prod_{j=1}^{m}t_j$.

Throughout this subsection, we set $t_i=1$ for all intermediate stages $i<m$ and assume $0<\theta_i<\pi/4$ for every register. This choice simplifies the recursive angle relation to a product of local overlap factors, allowing us to derive an explicit oracle-complexity bound. The recursive definition of $\gamma_m$ from Equation~\ref{equation:gamma_recursion} then expands to:
\begin{equation}
    \sin(2\gamma_m) = \prod_{i=1}^{m}\sin(2\theta_i),
    \quad \text{where} \quad
    \sin(\theta_i) = |\!\braket{x_i|\psi_i}\!|.
\end{equation}
Defining the global overlap angle $\theta$ by $\sin(\theta) = |\!\braket{x|\psi}\!| = \prod_{i=1}^{m}\sin(\theta_i)$ and using $\sin(2\theta_i) = 2\sin(\theta_i)\cos(\theta_i)$:
\begin{equation}
    \sin(2\gamma_m) = 2^m\sin(\theta)\prod_{i=1}^{m}\cos(\theta_i).
\end{equation}
Since $0<\theta_i<\pi/4$, we have $\prod_{i=1}^{m}2\cos(\theta_i)>1$ and hence $2\gamma_m\geq\theta$. Using the fact that $\sin(x)/x$ is decreasing for $0<x\leq\pi/2$, we have $\sin(2\gamma_m)/(2\gamma_m) \leq \sin(\theta)/\theta$, giving:
\begin{equation}
    \gamma_m \geq \frac{\theta}{2}\cdot\frac{\sin(2\gamma_m)}{\sin(\theta)}
    = 2^{m-1}\theta\prod_{i=1}^{m}\cos(\theta_i).
\end{equation}
Thus the magnitude of the rotation angle $\gamma_m$ relative to the initial overlap angle $\theta$ increases exponentially with the number of stages $m$, scaled by the dampening factor $\prod_{i=1}^{m}\cos(\theta_i)<1$. We choose the integer final-stage iteration count $t_m^*$ closest to the first maximum of $P(x_m)$:
\begin{equation}
    t_m^* = \left\lfloor\frac{\pi}{4|\gamma_m|}-\frac{1}{2}\right\rceil
    \leq \frac{\pi}{2^{m+1}\theta\prod_{i=1}^{m}\cos(\theta_i)},
    \label{eq:optimal_iteration}
\end{equation}
and the oracle cost of the final stage is
\begin{equation}
    C = 2^{m-1}t_m^*.
    \label{eq:round-query-cost}
\end{equation}
Comparing with the query complexity of quantum amplitude amplification $C_{\mathrm{QAA}} = \frac{\pi}{4\theta}$~\cite{AmplitudeAmplification}:
\begin{equation}
    \boxed{C \leq \frac{C_{\mathrm{QAA}}}{\prod_{i=1}^{m}\cos(\theta_i)}}.
    \label{eq:complexity}
\end{equation}
The overhead factor $1/\prod_{i=1}^{m}\cos(\theta_i)$ is determined entirely by the local overlaps within each register. To recover the remaining target components, we then apply the inverse iterates, whose cost can be reduced by cancelling neighbouring identical partial diffusion operators and repeated reflection blocks. To count the remaining oracle calls, define $X_0=S_x$ and $X_k=X_{k-1}S_{\psi_k}X_{k-1}$, so that $X_k$ contains $2^k$ oracle calls and satisfies $X_k^2=\mathbb{I}$. Since partial diffusions on different registers commute and square to the identity,
\[
    W_k=S_{\psi_k}\cdots S_{\psi_1}X_kS_{\psi_1}\cdots S_{\psi_k}.
\]
For two successive stages $k$ and $k-1$, cancelling the intervening partial diffusions brings the blocks $X_{k-2}$ and $X_{k-1}$ together. Their product simplifies as
\[
    X_{k-2}X_{k-1}
    =X_{k-2}\bigl(X_{k-2}S_{\psi_{k-1}}X_{k-2}\bigr)
    =S_{\psi_{k-1}}X_{k-2}.
\]
Thus, the two stages together contribute only the $2^{k-2}$ oracle calls in $X_{k-2}$, leaving the same recovery procedure on the remaining $k-2$ registers. With these cancellations, each gap between successive oracle calls of the full run contains exactly one partial diffusion operator. Furthermore, when $m$ is even, the full run ends with an oracle call, which can be removed. For two registers, the full run reduces, up to a global phase, to $(S_{\psi_2}S_xS_{\psi_1}S_x)^{t_2}\ket{\psi}$, which alternates the two partial diffusion operators between successive oracle calls and recovers the two-register circuit of~\cite{hardware} for single-target unstructured search. More generally, let $J_k$ denote the additional oracle calls required to recover $k$ registers after these cancellations. Then
\[
    J_k=J_{k-2}+2^{k-2},\qquad k\geq2,
    \qquad J_0=0,\quad J_1=1.
\]
The inverse iterates therefore add
\begin{equation}
     J_{m-1}=\frac{2^{m-1}-(-1)^{m-1}}{3}
    \label{eq:J}
\end{equation}
oracle calls. For even $m$ this is one fewer, as the final oracle call is removed, giving $\lfloor(2^{m-1}-1)/3\rfloor$ for every $m$. Since $t_m^*\geq1$, this is no greater than the final-stage cost $2^{m-1}t_m^*$. Applying Equation~\ref{eq:complexity} therefore gives
\begin{equation}
    C_{\mathrm{total}}
    =C+\left\lfloor\frac{2^{m-1}-1}{3}\right\rfloor
    \leq2C
    \leq\frac{2C_{\mathrm{QAA}}}{\prod_{i=1}^{m}\cos(\theta_i)}.
    \label{eq:total-complexity}
\end{equation}
Thus,
\begin{equation}
    C_{\mathrm{total}} = O\!\left(\frac{C_{\mathrm{QAA}}}{\prod_{i=1}^{m}\cos(\theta_i)}\right),
\end{equation}
an overhead of $O(1/\prod_{i=1}^{m}\cos(\theta_i))$ oracle calls per instance relative to amplitude amplification on the same problem.

This overhead describes a single instance of the algorithm, which succeeds with probability $P^*(x)$. To obtain the expected oracle cost of finding a target, we also have to account for repetition given unsuccessful outcomes. We assume that measured outcomes can be verified as target outcomes, classically in the case of unstructured search, and exclude this verification cost from the quantum resource counts. The expected number of instances required to obtain a target outcome is $1/P^*(x)$, giving
\begin{equation}
    \mathbb{E}[Q]
    =\frac{C_{\mathrm{total}}}{P^*(x)}
    \leq\frac{2C_{\mathrm{QAA}}}
    {P^*(x)\prod_{i=1}^{m}\cos(\theta_i)}.
    \label{eq:expected-queries}
\end{equation}
The inequality follows from Equation~\ref{eq:total-complexity}. When $P^*(x)\approx\prod_i\cos^2(\theta_i)$, the upper bound is approximately $2C_{\mathrm{QAA}}/\prod_i\cos^3(\theta_i)$. As established in Subsection~\ref{sec:overall-succ}, if the local success probabilities $p_i=\sin^2(\theta_i)$ satisfy $\sum_i p_i\leq\kappa$ for a fixed constant $\kappa$, then
\[
    \frac{1}{\prod_{i=1}^{m}\cos(\theta_i)}\leq2^\kappa.
\]
Under the same bounded-sum condition, $p=\prod_i p_i\to0$ implies that the smallest local probability $\min p_i\to0$. With $t_i=1$, Equation~\ref{equation:gamma_recursion} gives $0<\gamma_m\leq\min \theta_i$, so the residual correction vanishes. Equation~\ref{eq:overall-succ} therefore bounds the full success probability away from zero, giving $\mathbb{E}[Q]=O(C_{\mathrm{QAA}})$ by Equation~\ref{eq:expected-queries}. The quadratic query scaling is thus retained when recovering the full target with any fixed success probability less than one.
\subsection{Non-Oracle Overhead}

Having established the oracle complexity, we now consider the cost of the partial diffusion operators. We count their applications after cancellations, then compare the expected state preparation sub-circuit counts with those of standard amplitude amplification. As in the oracle-complexity analysis, we set $t_i=1$ at all intermediate stages.

Recall that the recursive operator is defined as
\begin{equation}
    W_i = (S_{\psi_i} W_{i-1})^{t_i} \, S_{\psi_i} \, (W_{i-1} S_{\psi_i})^{t_i}, \quad W_0 = S_x.
\end{equation}
Using the decomposition of $W_i$ from Subsection~\ref{subsec:oracle}, the central block $X_i$ contains $2^{i-k}$ applications of $S_{\psi_k}$, while the surrounding partial diffusions contribute two more. The resulting diffusion count is
\begin{equation}
    (2^0 + 2)S_{\psi_i} + (2^1 + 2)S_{\psi_{i-1}} + (2^2 + 2)S_{\psi_{i-2}} + \cdots + (2^{i-1} + 2)S_{\psi_1}.
\end{equation}
After the further cancellations described in Subsection~\ref{subsec:oracle} and omission of the initial partial diffusions that contribute only a global phase, the final-stage amplification contributes $t_m$ applications of $S_{\psi_m}$ and $2^{m-1-i}t_m$ applications of each $S_{\psi_i}$ for $i<m$. Resolving the remaining registers adds $J_{m-1-i}$ applications of each $S_{\psi_i}$ for $i<m$, where $J_k=(2^k-(-1)^k)/3$.

Given $A=\bigotimes_{i=1}^{m}A_i$, each partial diffusion $S_{\psi_i}=A_i(I-2\ket{0_i}\bra{0_i})A_i^\dagger$ requires one application each of $A_i$ and $A_i^\dagger$, together with a local reflection about $\ket{0_i}$. Counting applications of $A_i$ and $A_i^\dagger$ together, and taking $t_m=t_m^*$, the total sub-circuit count within the diffusion operators is
\begin{equation}
\begin{split}
    G_{\mathrm{recursive}}
    &=2t_m^*A_m
    +2\sum_{i=1}^{m-1}
    \bigl(2^{m-1-i}t_m^*+J_{m-1-i}\bigr)A_i\\
      &\leq\frac{\pi}{2\theta\prod_{j=1}^{m}\cos(\theta_j)}
    \left[
        \frac{1}{2^{m-1}}A_m
        +2\sum_{i=1}^{m-1}\frac{1}{2^i}A_i
    \right].
\end{split}
\label{eq:non-oracle-cost}
\end{equation}
Where resolving the lower registers adds no more inner-stage diffusion applications than the final-stage amplification itself, increasing these counts by at most a factor of two. In QAA, each global diffusion applies every sub-circuit once in each direction, giving
\begin{equation}
    G_{\mathrm{QAA}}
    =2t^*\sum_{i=1}^{m}A_i
    \leq\frac{\pi}{2\theta}\sum_{i=1}^{m}A_i.
\end{equation}
Let $R_i$ denote the number of applications of $A_i$ or its inverse in QAA divided by the corresponding number in the construction in this work, counting only the diffusion operators within a single instance. Thus, $R_i>1$ favours the recursive construction, while $R_i<1$ favours QAA. Since integer rounding gives $t^*\geq\pi/(8\theta)$, Equation~\ref{eq:non-oracle-cost} yields
\begin{equation}
    R_i\geq2^{i-2}\prod_{j=1}^{m}\cos(\theta_j),
    \qquad 1\leq i\leq m.
\end{equation}

To account for repetition, we divide each count by the corresponding success probability. The ratio of expected application counts therefore satisfies
\begin{equation}
\begin{split}
    R_i\,\frac{P^*(x)}{P_{\mathrm{QAA}}(x)}
    &\geq
    \frac{2^{i-2}P^*(x)}{P_{\mathrm{QAA}}(x)}
    \prod_{j=1}^{m}\cos(\theta_j)\\
    &\approx
    \frac{2^{i-2}}{P_{\mathrm{QAA}}(x)}
    \prod_{j=1}^{m}\cos^3(\theta_j),
    \qquad 1\leq i\leq m,
\end{split}
\end{equation}
where integer rounding gives $\cos^2(\theta)\leq P_{\mathrm{QAA}}(x)\leq1$. For small local overlap angles $\theta_i$, with which the global angle $\theta$ is also small, $P_{\mathrm{QAA}}(x)\approx1$ and $P^*(x)\approx\prod_j\cos^2(\theta_j)$. The lower bound is then approximately $2^{i-2}\prod_j\cos^3(\theta_j)$.

Whenever ratio exceeds one, the construction uses fewer expected applications of $A_i$ or its inverse than standard amplitude amplification. This becomes more favourable at higher stages, with the factor increasing exponentially with the partition index. This favours placing cheaper sub-circuits at the lower stages and more expensive sub-circuits at the higher stages, so that the most costly operations are repeated fewer times. Further savings may arise from the reflections about $\ket{0}$, which are omitted from these sub-circuit counts. In the recursive scheme these are local reflections on $n_i$ qubits, each requiring a multi-controlled gate of size $n_i$. In standard QAA the reflection acts on all $n$ qubits, requiring a multi-controlled gate of size $n$. Together, these savings can give the recursive scheme a lower expected non-oracle gate count than standard amplitude amplification, accounting for the success probability of each approach, for problems satisfying the state preparation and marked-subspace factorisation assumptions.
\subsection{Application to Unstructured Search of a Single Target}
\label{subsec:unstructured}

As an example, we apply the algorithm to unstructured search for a single target, a problem that naturally satisfies the tensor decomposition constraints in Equation~\ref{eq:decomp}. Here $A=H^{\otimes n}=\bigotimes_{i=1}^{m}H^{\otimes n_i}$, and the marked computational-basis state factorises as $\ket{x}=\bigotimes_{i=1}^{m}\ket{x_i}$, so its projector satisfies $P_G=\bigotimes_{i=1}^{m}\ket{x_i}\bra{x_i}$. We partition the $n=\log_2 N$ search qubits into $m$ registers of equal size $s=n/m$. Equal register sizes are not required by the construction, but provide a baseline for analysing its resource requirements. As in Subsection~\ref{subsec:oracle}, we set $t_i=1$ at all intermediate stages and choose $t_m=t_m^*$. Taking $s\geq2$ gives the local overlap angles $\theta_i=\arcsin(2^{-s/2})<\pi/4$. We consider $m\geq2$; for $m=1$, the construction reduces to standard Grover search.

For equal registers, the sum of the local success probabilities is $\sum_i p_i=n/(s2^s)$, and the cosine product in Equations~\ref{eq:complexity} and~\ref{eq:overall-succ} becomes
\begin{equation}
    \prod_{i=1}^{m}\cos^2(\theta_i)
    =(1-2^{-s})^{n/s}
    \approx\exp\!\left(-\frac{n}{s\,2^s}\right).
    \label{eq:unstructured-product}
\end{equation}
This sum remains bounded when $s2^s=\Omega(n)$. For $s\geq\log_2 n$ it is at most $1/s$ and tends to zero as $n$ grows. By Subsections~\ref{sec:overall-succ} and~\ref{subsec:oracle}, the success probability then tends to one and the final-stage oracle overhead converges to unity, while the $J_{m-1}\leq N^{1/s}$ oracle calls needed to recover the remaining registers are negligible compared with $\sqrt{N}$. Thus, registers of at least $\log_2 n=\log_2(\log_2 N)$ qubits retain the $O(\sqrt{N})$ expected oracle complexity after accounting for recovery of the remaining registers and repetition after unsuccessful outcomes. For a fixed register size, this sum instead grows linearly with $n$, so the oracle overhead and the repetition required after unsuccessful outcomes eventually grow with the problem size.

For the non-oracle cost, let $G(s)$ denote the cost of a complete diffusion operator on $s$ qubits, including the Hadamard gates and the local reflection about $\ket{0}$. The application counts underlying Equation~\ref{eq:non-oracle-cost} sum to at most $C_{\mathrm{total}}$ partial diffusion operators per instance, so the expected diffusion cost is at most
\begin{equation}
    \frac{G(s)\,C_{\mathrm{total}}}{P^*(x)}
    =G(s)\,\mathbb{E}[Q]
    =O\!\left(G(s)\sqrt{N}\right)
    \label{eq:grover-non-oracle}
\end{equation}
for $s\geq\log_2 n$, against $O\!\left(G(n)\sqrt{N}\right)$ for standard Grover search. The construction therefore replaces the global diffusion cost $G(n)$ with the local cost $G(s)$ while retaining the quadratic expected query scaling. Choosing $s=\lceil\log_2 n\rceil$ and adding the remaining $n-s\lfloor n/s\rfloor$ qubits to one register gives registers of $s$ to $2s-1$ qubits and confines each partial diffusion to $O(\log n)=O(\log\log N)$ qubits. Assuming a diffusion operator has a cost linear in the number of qubits it acts on, $G(s)=O(s)$ and the expected non-oracle gate count is $O(\sqrt{N}\log\log N)$. Under the same assumption, this matches~\cite{grover_trade-offs_2002} and is a $\log\log N$ factor above the $\Theta(\sqrt{N})$ of~\cite{hardwareGrover2}.

Non-uniform register sizes may be better suited to the hardware topology. The same asymptotic expected query complexity is retained if each register satisfies $n_i\geq\log_2(\log_2 N)$. In this case,
\begin{equation}
    \sum_{i=1}^{m}p_i
    =\sum_{i=1}^{m}2^{-n_i}
    \leq\frac{m}{n}
    \leq\frac{1}{\log_2 n}
    \longrightarrow0,
\end{equation}
so the success probability again tends to one, the final-stage oracle overhead converges to unity, and the recovery calls remain negligible relative to $\sqrt{N}$.

The analysis is not restricted to a single target. For a set $G$ of marked strings, the factorisation in Equation~\ref{eq:decomp} holds exactly when $G$ is a Cartesian product $G=G_1\times\cdots\times G_m$ of sets of marked values on the individual registers, so that $P_G=\bigotimes_{i=1}^{m}P_{G_i}$ with $P_{G_i}=\sum_{x_i\in G_i}\ket{x_i}\bra{x_i}$. The local overlap angles then become $\theta_i=\arcsin\sqrt{|G_i|/2^{n_i}}$, and the results of this section apply provided $|G_i|<2^{n_i-1}$ for every register, which ensures $\theta_i<\pi/4$. Under the bounded-sum condition of Subsection~\ref{subsec:oracle}, the expected oracle cost is then $O(\sqrt{N/|G|})$. A general set of marked strings does not have this form and falls outside the present analysis.

\section{Evaluation}
\label{sec:evaluation}
The preceding section establishes the theoretical foundations of our approach: the recursive construction built from local reflections, the observed structure of the principal angles between such reflections, and exact expressions for the algorithm's dynamics and oracle overhead. In this section we focus on evaluating these results in practice, applying the recursive search scheme to unstructured search for a single target state. We simulate the algorithm on an 18-qubit search instance, instantiating the scheme for search instances divided up into two and three registers of equal size, showing results consistent with the equations in Subsection~\ref{sec:succ} and Appendix~\ref{app:run}, and quantifying the resulting circuit depth reductions. We then evaluate these expressions, together with individually transpiled diffusion operators, for four partitioning strategies over $n=20$ to $256$ qubits, and characterise the regimes in which our scheme offers valuable depth and gate count advantages over Grover search.

\subsection{Circuit Simulation}

We first illustrate the scheme through noiseless circuit simulation on an 18-qubit unstructured search problem, examining both measurement outcomes and circuit depth against Grover search. An $18$-qubit search instance was chosen as it could be partitioned uniformly into either two registers (of $9$ qubits), or three  registers (of $6$ qubits), while remaining classically simulable.

\begin{figure}[t]
\begin{center}
\includegraphics[width=1\textwidth]{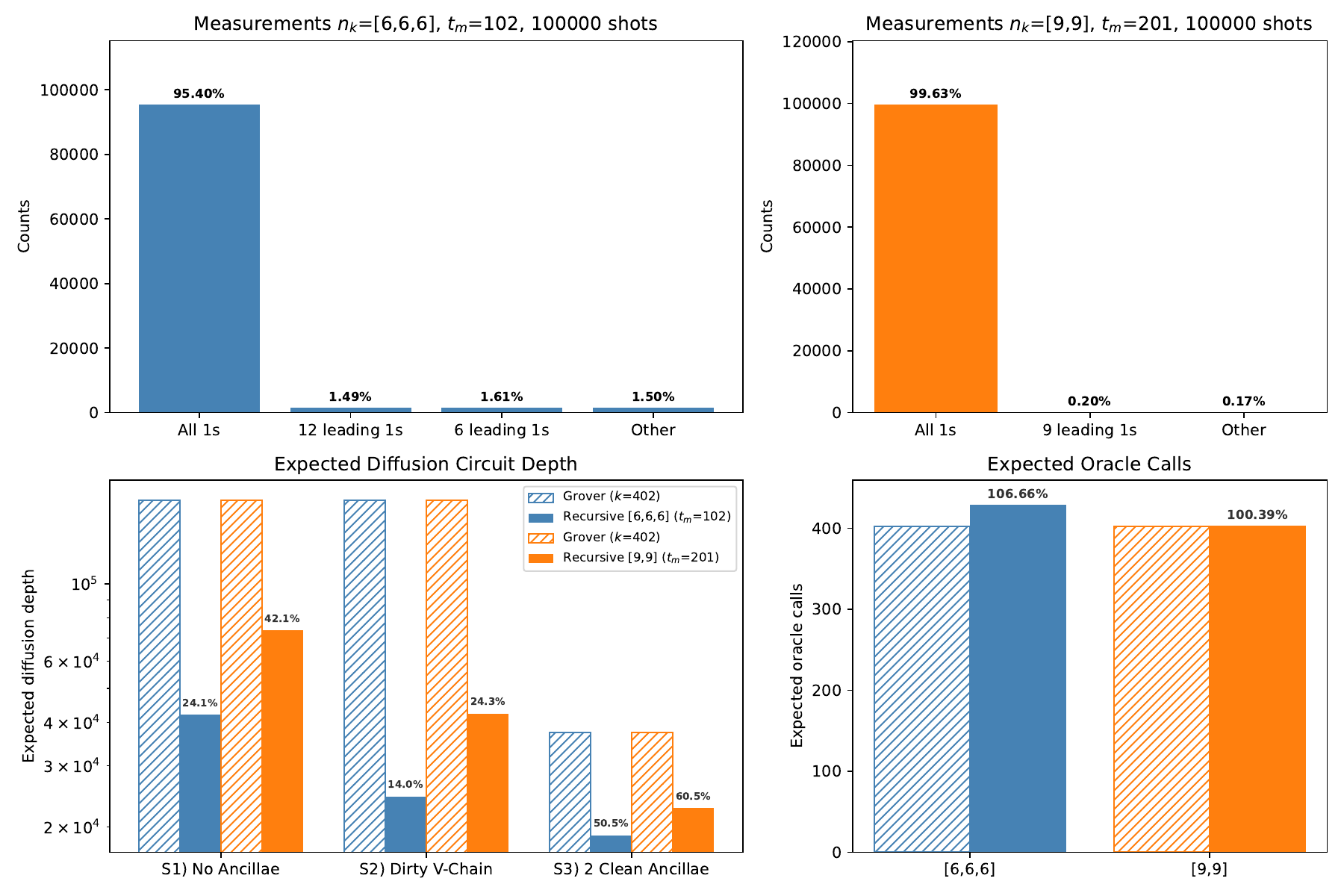}
\caption{Simulation and depth comparison of the recursive search scheme against standard Grover search on an 18-qubit single-target problem. \textbf{Top row:} Measurement outcome distributions for the 3-stage $[6,6,6]$ (left) and 2-stage $[9,9]$ (right) configurations over $100\,000$ shots, showing the proportion of shots returning the correct target bitstring and partial matches. \textbf{Bottom left:} Expected circuit depth of just the diffusion operators (log scale) for both configurations under three decomposition scenarios of the MCX gate in the diffusion operator. S1: no ancillae available for any construction, S2:  idle qubits from other partitions could be used to form a dirty V-chain, S3: two clean ancillae available for each construction. The circuits were transpiled onto a fully connected topology with CX as the native two-qubit gate. Percentages give the recursive scheme's depth for implementing all of the diffusion steps as a percentage of that of Grover search, so that a label of $24.1\%$ corresponds to a depth reduction of $75.9\%$. \textbf{Bottom right:} Expected oracle call counts, with percentages showing the recursive scheme's calls relative to the Grover search. }
\label{fig:verification-results}
\end{center}
\end{figure}

\subsubsection{Measurement Outcomes}
The algorithm was implemented in Qiskit and evaluated for an 18-qubit search of the computational state $\ket{1}^{\otimes 18}$. Two stage configurations were tested (a 3-stage configuration with 6 qubits per stage, and a 2-stage configuration with 9 qubits per stage). We set $t_m=t_m^*$, the integer nearest the first success-probability maximum of Equation~\ref{eq:stage-prob}, and $t_i = 1$ for the intermediate iterates, yielding $t_m=102$ and $201$ respectively. Each  configuration was run for $100\,000$ shots, giving standard errors of at most $0.07$ percentage points on the overall success and failure rates below. The extracted measurement probabilities are shown in the top row of Figure~\ref{fig:verification-results}.

The simulation results agree with the theoretical predictions. The 3-stage configuration measured the target bitstring $95.40\%$ of the time, against a theoretical expectation of $95.39\%$ (Appendix~\ref{app:run}) and a lower bound of $94.64\%$ from Equation~\ref{eq:overall-succ}. The 2-stage configuration observed $99.63\%$ success, against a theoretical value of $99.61\%$ (Appendix~\ref{app:run}) and a bound of $99.47\%$. We note that the attainable success probability remains slightly below that of Grover search, which would exceed $99.99\%$ on the same problem size. The 2-stage configuration achieves a higher success probability than the 3-stage configuration, since the ceiling of the success probability decreases both as the number of stages grows and individual stage sizes shrink.

Importantly, the failure cases also reveal useful structure. In the 3-stage configuration, $4.60\%$ of shots did not return the correct result, but of these $67.39\%$ still contained the correct $x_3$. Of those with a correct $x_3$, $48.00\%$ also had the correct $x_2$. This is expected as each successive stage's success is conditional on the state of the previous stages. For the 2-stage configuration, $53.21\%$ of the $0.37\%$ of failures still contained the correct $x_2$. So when only a prefix of the target is required, the effective success probability increases. As per Equation~\ref{eq:app-prefix} for the $3$-stage configuration, the success probability rises to $96.90\%$ if only the first $12$ bits of the target are needed, and to $98.44\%$ for the first $6$ bits, against $95.39\%$ for the full target (Equations~\ref{eq:app-base} and~\ref{eq:app-prefix}).

\subsubsection{Depth Comparison}
\label{subsec:sim-depth}

We next assess the circuit depth of the simulated implementations. Both configurations were compared against Grover search for a single target. Both schemes use the iteration counts corresponding to their first success probability maximum: $t_m=t_m^*$, and $k=402$ for Grover. Expected costs divide the cost of one run by its success probability (Equation~\ref{eq:expected-queries}), which is $99.9998\%$ for Grover and $95.39\%$ and $99.61\%$ for the 3- and 2-stage configurations, assuming that the whole circuit is rerun after an unsuccessful classical verification, which would occur in $0.39\%$--$4.61\%$ of these runs. We note that we could instead under-amplify each scheme to further reduce their expected oracle costs as per~\cite{GroverMultiple, zalkaOptimal}; however, this would change the relative oracle overheads and circuit depths reported in this section by less than one percentage point, and does not take into account the added cost of the additional setup and measurements required.

The cost of implementing the diffusion operators depends heavily on how the multi-controlled X (MCX) gates they contain are decomposed, which in turn depends on the topology of the underlying hardware and the available ancilla resources. To capture this dependence, we evaluated three decomposition scenarios for the MCX gates, varying the number of available ancillae while fixing the connectivity to a fully connected topology. In S1, no ancilla qubits were permitted and each local diffusion operator could access only the qubits within its own group, illustrating the circuit depth that results when no inter-partition communication is allowed. Here Qiskit's default synthesis applies the linear-depth construction of~\cite{dasilva2022}. In S2, idle qubits from other groups were made available as dirty ancillae for V-chain decomposition~\cite{mcx1,mcx2}, enabling more efficient transpilation of the partial diffusion operator without requiring any dedicated ancilla qubits. In S3, two clean ancillae were provided to both schemes to demonstrate the comparison under a more-efficient decomposition for the global diffusion operator. At two clean ancillae, Qiskit's default synthesis applies the logarithmic-depth construction of~\cite{khattar2025rise} and adding additional ancillae would leave the transpiled circuits unchanged. Here the two ancillae are shared by all groups, although assigning a separate pair to each group would produce circuits with the same depth and gate count. All circuits were transpiled with CX as the native two-qubit gate. We used the gate set $\{u,\mathrm{CX}\}$, Qiskit's in-built transpilation optimiser at level~$3$ and a fixed transpiler seed. We compared both the expected circuit depth of the diffusion steps and the expected number of oracle calls. The diffusion circuit depth of each scheme was calculated by transpiling the full circuit with a barrier in place of each oracle call. These results are shown in the bottom row of Figure~\ref{fig:verification-results}.

Both stage configurations require significantly shallower circuits to implement the diffusion steps relative to Grover search under all three scenarios, with a small oracle overhead. As expected, the depth reductions are most pronounced when ancilla qubits are unavailable and the tested MCX decompositions are least efficient. Under S1, in which none of the constructions were permitted any ancillae, the 3-stage and 2-stage configurations reduce the circuit depth for diffusion by $75.91\%$ and $57.90\%$ respectively. Under S2, where the recursive schemes were able to use the idle qubits from other partitioned registers, these reductions relative to Grover increase to $85.99\%$ and $75.71\%$. Furthermore, the constructions only require $6.66\%$ and $0.39\%$ additional oracle calls relative to Grover, for the 3-stage and 2-stage configurations respectively. When two clean ancillae are introduced in S3 --- which allows for a more efficient MCX decomposition scheme --- the depth advantage narrows but remains significant, with the 3-stage configuration reducing the diffusion circuit depth by $49.54\%$ and the 2-stage by $39.55\%$.

From these results, we can characterise the regime in which depth savings persist even when the circuit implementation of the oracle is significantly deeper than that of the global diffusion operator. Within this regime, the depth reduction from the partial diffusion operators outweighs the additional depth incurred by the extra oracle calls, yielding a net reduction in overall circuit depth. Let $d_O$ and $d_G$ be the circuit depths of the oracle and the global diffusion operator, and let $\mathbb{E}[D]$ and $\mathbb{E}[Q]$ be the expected non-oracle depth and oracle calls of a scheme. The recursive scheme (R) and Grover search (G) have equal expected total depth $\mathbb{E}[D]+d_O\,\mathbb{E}[Q]$ at the break-even ratio\begin{equation}
    r^*=\frac{d_O}{d_G}
    =\frac{\mathbb{E}[D_{\mathrm{G}}]-\mathbb{E}[D_{\mathrm{R}}]}
    {d_G\bigl(\mathbb{E}[Q_{\mathrm{R}}]-\mathbb{E}[Q_{\mathrm{G}}]\bigr)},
    \label{eq:break-even}
\end{equation}below which the recursive scheme is shallower. Here we treat the oracle and diffusion operators as individual entities and do not take into account any further optimisations over the entire circuit. For the 3-stage configuration, the recursive scheme remains shallower overall provided the oracle depth does not exceed approximately $11.4$, $12.9$, or $7.4$ times the global diffusion operator depth for S1, S2, and S3 respectively. For the 2-stage configuration, these thresholds are higher, at $147.9$, $193.4$, and $101.0$ times. With two clean ancillae, for an $18$-qubit unstructured single-target search and an oracle with the same circuit depth as the global diffusion operator, the 2- and 3-stage configurations have a $19.58\%$ and $21.44\%$ lower expected total depth than Grover. With an oracle $10\times$ as deep, the 2-stage configuration remains $3.24\%$ lower, while the 3-stage configuration is $1.55\%$ higher. At this problem size, more, smaller stages therefore give the lower expected total depth when the oracle is relatively cheap to implement, and fewer, larger stages when it is relatively expensive. We explore this configuration choice further in Subsection~\ref{subsec:depth-savings}. 

The expected costs above divide the cost of one run by its success probability. If the goal is instead to reach a minimum target success probability, say $p^*\geq0.99$, the comparison changes, and more strongly for configurations whose success probability falls below the target. In this scenario, each scheme would have to be run up to $R$ times, for the smallest $R$ with $1-(1-P)^R\geq p^*$, where $P$ is the success probability of one run at an iteration count chosen to minimise the oracle calls. For an $18$-qubit single target, Grover search would need $k=376$ iterations in a single run ($P=0.990$). The $2$-stage configuration, whose success probability at $t_m^*$ is $0.996$, would also need a single run, at $t_m=192$ ($P=0.991$) and $384$ oracle calls, $2.1\%$ more than Grover. The $3$-stage configuration reaches only $0.954$ at $t_m^*$, below the target, so it would need two runs ($R=2$), at $t_m=87$ ($P=0.901$) and up to $698$ oracle calls, $86\%$ more than Grover. A minimum target success probability therefore favours fewer, larger stages, whose success probability in a single run can exceed the target.

\subsection{Analysis on Larger Search Spaces}
We now examine how the scheme scales to larger search sizes. Using the previously-derived expressions from Section~\ref{sec:algorithm} and Appendix~\ref{app:run}, we explore the failure probability and oracle overhead over $n = 20$ to $n = 256$ qubits. These expressions were evaluated in arbitrary-precision arithmetic to remain accurate at large $n$. Combining these expressions with individually transpiled diffusion operators, we also estimate the circuit depth and CX count relative to Grover search.

We investigate four strategies for partitioning the search space into stages of uniform size $s$: a fixed number of stages, $m=4$, with $s=n/4$; a stage size that grows as the square root of the qubit count, $s=\lceil\sqrt{n}\rceil$; the stage size $s=\lceil\log_2 n\rceil$ of Subsection~\ref{subsec:unstructured}; and a fixed stage size, $s=7$. Only values of $n$ that divide into equal stages are shown.

\begin{figure}[t]
\begin{center}
\includegraphics[width=1\textwidth]{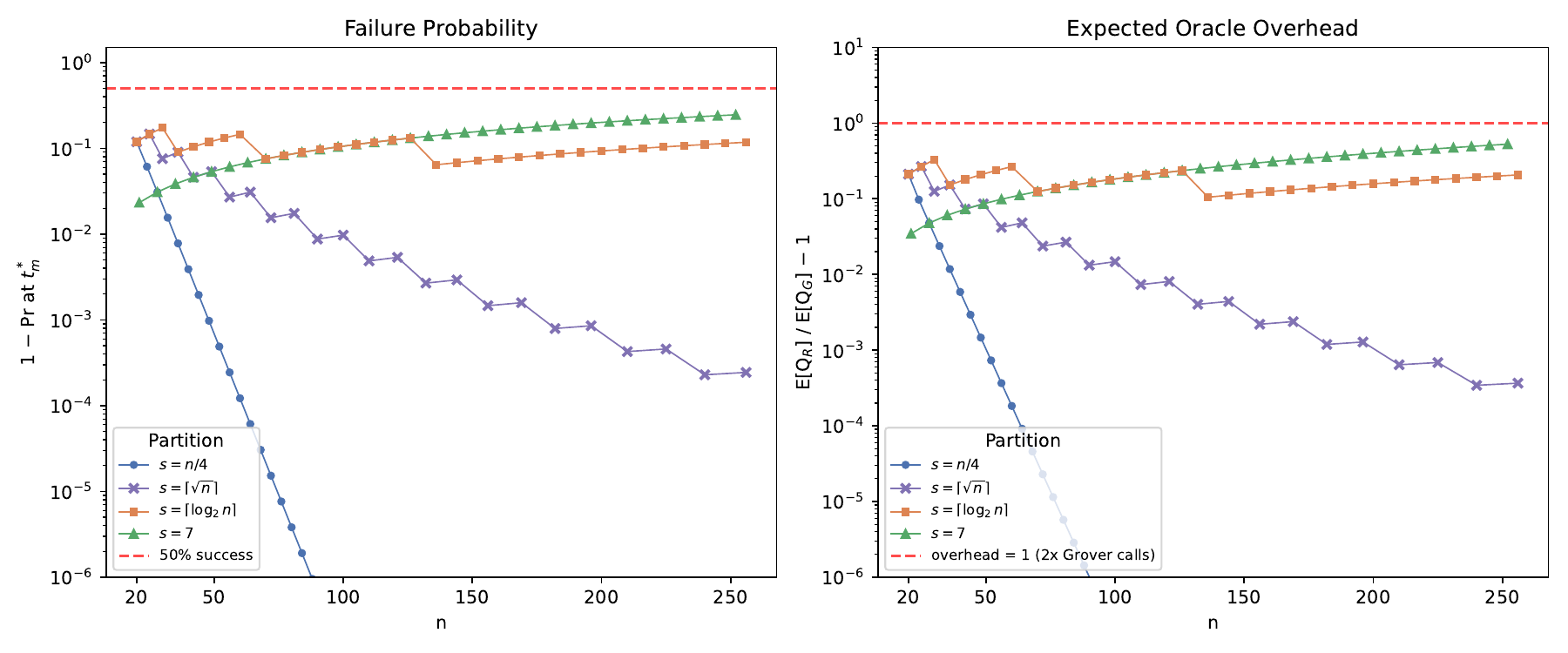}
\caption{Scaling behaviour of the recursive search scheme for each of the four partition strategies: $s = n/4$, $s=\lceil \sqrt{n} \rceil$, $s=\lceil \log_2{n} \rceil$, and $s = 7$. These were computed using the expressions from Subsection~\ref{sec:overall-succ}, Subsection~\ref{subsec:oracle} and Appendix~\ref{app:run} applied to unstructured searches over $n = 20$ to $n = 256$ qubits with a single target state. Only values of $n$ that divide into equal stages are shown. \textbf{Left:} Failure probability $1-P^*(x)$ per run at $t_m^*$. \textbf{Right:} Relative oracle overhead, defined as the fractional increase in expected oracle calls over standard Grover search at its first success maximum. For visual clarity, both vertical axes are cut at $10^{-6}$, below which $s=n/4$ falls from about $n=90$.}

\label{fig:scaling-properties}
\end{center}
\end{figure}

\subsubsection{Failure Probability and Oracle Overhead}

Figure~\ref{fig:scaling-properties} shows the failure probability and relative oracle overhead for the four partitioning strategies, on a single-target search over $n = 20$ to $n = 256$ qubits.  As shown in Subsections~\ref{sec:overall-succ} and~\ref{subsec:oracle}, both the failure probability $1-P^*(x)$ and the oracle overhead are governed by $\prod_i(1-p_i)$, which for uniform stage sizes is $(1-2^{-s})^{n/s}\approx\exp\!\bigl(-n/(s2^s)\bigr)$. When the stage size grows polynomially with $n$, as for $s=n/4$ and $s=\lceil\sqrt{n}\rceil$, the failure probability decreases exponentially with $s$, reaching $9.31\times10^{-10}$ at $n=128$ for $s=n/4$ and $2.44\times10^{-4}$ at $n=256$ for $s=\lceil\sqrt{n}\rceil$. When the stage size grows logarithmically, as for $s=\lceil\log_2 n\rceil$, $2^s\geq n$ gives $n/(s2^s)\leq1/s$, so the failure probability remains below approximately $1/s$. Over the plotted range of $n$ it lies between $0.0644$ and $0.1734$, rising while $s$ is constant and falling each time $s$ increases. The oracle overhead follows a similar pattern, falling between $10.5\%$ and $33.0\%$ more oracle calls than Grover search. For a fixed stage size, as for $s=7$, $n/(s2^s)$ instead grows linearly with $n$, and the failure probability exceeds $1/s=0.143$ around $n=2^s=128$ ($0.1384$ at $n=133$ and $0.1452$ at $n=140$). The failure probability and oracle overhead for $s=7$ are the lowest of the plotted strategies at small search sizes, at $0.0233$ and $3.46\%$ for $n=21$ (where $s=7=n/3$), but rise to $0.2460$ and $52.7\%$ at $n=252$. With $s=7$, the failure probability would eventually exceed $1/2$ from $n\approx s2^s\ln2\approx620$. Note that for $n=65$ to $128$, $\lceil\log_2 n\rceil=7$, so the logarithmic and fixed strategies give the same partition and their points coincide. Other strategies coincide at individual sizes for the same reason, as $s=n/4$, $s=\lceil\sqrt{n}\rceil$ and $s=\lceil\log_2 n\rceil$ do at $n=20$, where each gives four stages of five qubits. These values illustrate the analysis of Subsection~\ref{subsec:unstructured}, where for unstructured search with a single target, fewer, larger stages give both a lower failure probability and a lower oracle overhead. Both remain bounded as $n$ increases provided the stage size is at least $\log_2 n$, and vanish rapidly when it grows polynomially, whereas for a fixed stage size they eventually grow with $n$.

\subsubsection{Circuit Depth Savings}\label{subsec:depth-savings}

\begin{figure}[t]
\begin{center}
\includegraphics[width=1\textwidth]{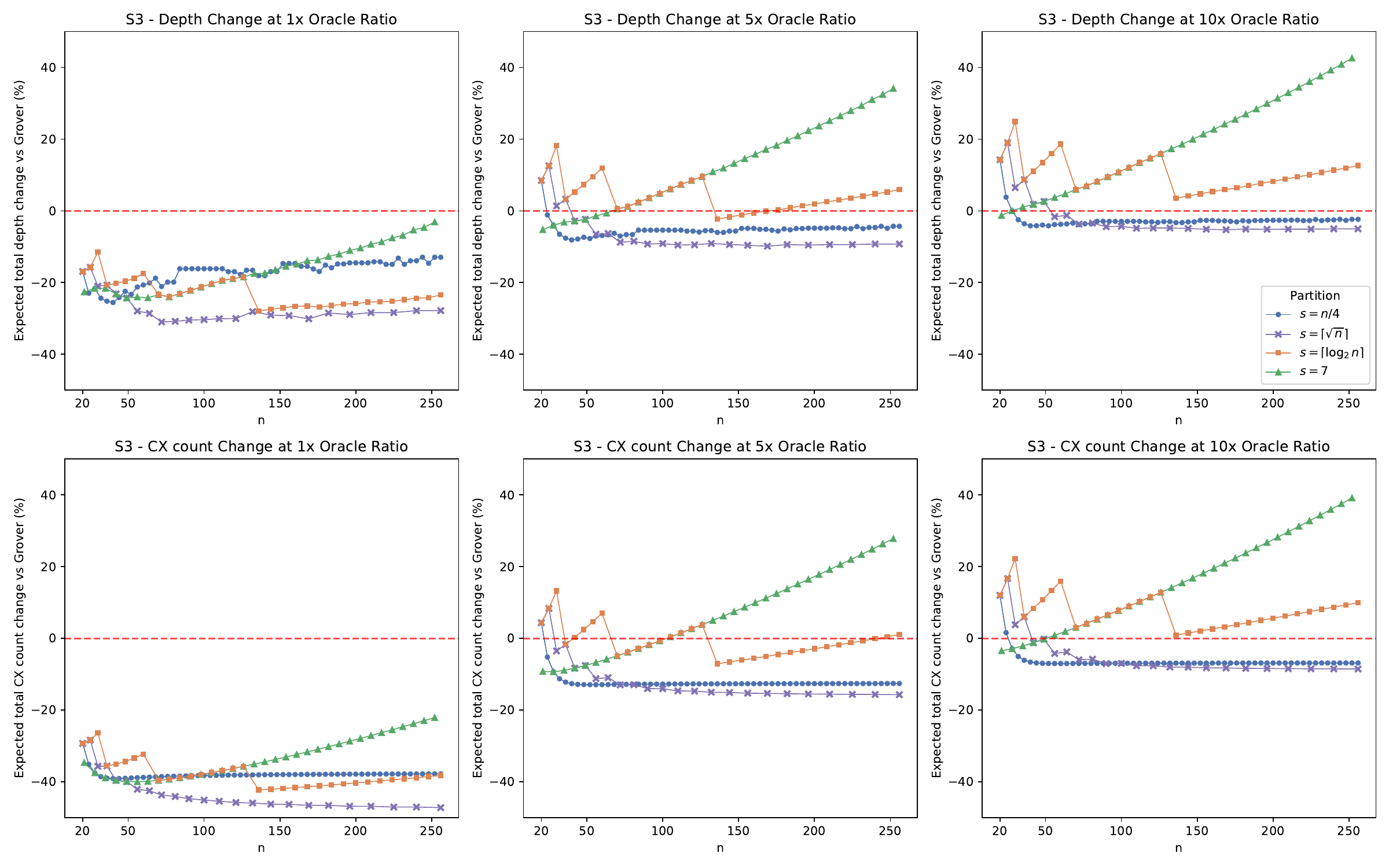}
\caption{Change in expected total circuit depth and CX gate count of the recursive search scheme relative to Grover search, for the four partitioning strategies of Figure~\ref{fig:scaling-properties} over $n = 20$ to $n = 256$ qubits, with $t_m=t_m^*$ and Grover at its first success maximum. All values use S3 and are estimated from the exact expressions and individually transpiled diffusion operators. \textbf{Top row:} Change in expected total circuit depth, with a single oracle call contributing $1\times$, $5\times$ and $10\times$ the depth of the global diffusion operator ($r=1$, $5$ and $10$) in the left, middle and right columns. \textbf{Bottom row:} The same for the CX gate count, with the oracle's CX count taken at the same ratio. The dashed red line marks parity with Grover search, and values below it indicate an advantage for the recursive scheme. Only values of $n$ that divide into equal stages are shown. The jumps come from the integer stage sizes and, for depth, from the MCX decomposition, whose depth does not increase monotonically with the number of control qubits.}

\label{fig:depth-scaling}
\end{center}
\end{figure}

Given the assessed oracle overhead of each partitioning strategy, we now estimate the resulting circuit depth and CX count relative to Grover search, over the same range of $n$ and for the same four strategies. All circuits use the S3 setting of Subsection~\ref{subsec:sim-depth}, with two clean ancillae on a fully connected topology and each MCX gate decomposed with the logarithmic-depth construction of~\cite{khattar2025rise}. We note that this scenario is not necessarily representative of near-term hardware, as at $n=256$ it assumes $258$ fully connected qubits. Nevertheless, it provides a conservative setting in which to assess the recursive scheme, since the relative depth difference between a partial diffusion operator and the global one is smaller under its decomposition than under those of S1 or S2, which favours Grover search. For each $n$, we transpiled the global diffusion operator and the partial diffusion operators of each stage size separately, and multiplied the depth and CX count of each by its number of applications as per Equation~\ref{eq:non-oracle-cost}. The decomposition remains tractable at large $n$ as the transpilation acts on the gate-level description of a circuit and does not simulate any actual state space. At $n=256$, the global diffusion operator transpiles onto 258 qubits at a depth of $201$ and requires only $1524$ CX gates. The cost of an oracle call is modelled as a multiple $r$ of that of the global diffusion operator, as in Equation~\ref{eq:break-even}, with the same ratio used for depth and for CX count. The expected total cost is the sum of the oracle and diffusion contributions divided by the success probability per run, and Figure~\ref{fig:depth-scaling} shows its change relative to Grover search for $r=1$, $5$ and $10$.

The top row of Figure~\ref{fig:depth-scaling} shows the change in expected total depth under this model for each strategy. With an oracle the same depth as the global diffusion operator ($r=1$), all four strategies give shallower circuits than Grover search at the largest sizes plotted. At $n=256$ the depth is reduced by $12.94\%$ for $s=n/4$, $27.83\%$ for $s=\lceil\sqrt{n}\rceil$ and $23.46\%$ for $s=\lceil\log_2 n\rceil$, and at $n=252$ by $3.12\%$ for $s=7$. As the oracle becomes deeper, the reduction narrows, and it persists only for the stage sizes that grow polynomially with $n$. At $r=5$ and $r=10$, the reductions at $n=256$ are $4.31\%$ and $2.35\%$ for $s=n/4$, and $9.25\%$ and $5.03\%$ for $s=\lceil\sqrt{n}\rceil$. With $s=\lceil\log_2 n\rceil$ the circuit is instead $5.96\%$ and $12.64\%$ deeper than Grover search, and with $s=7$ it is $34.12\%$ and $42.58\%$ deeper at $n=252$. These trends broadly follow those of the previous subsection. However, despite their higher failure probability and oracle overhead, the $s=\lceil\sqrt{n}\rceil$ and $s=\lceil\log_2 n\rceil$ stage configurations yield shallower circuits than $s=n/4$ at $r=1$, from $n=56$ and $n=70$ respectively. For $s=\lceil\sqrt{n}\rceil$ this advantage persists at the deeper oracle ratios $r=5$ and $r=10$, from $n=72$ and $n=100$, where it gives the shallowest circuits of the four strategies. The depth saved by additional smaller diffusion operators can outweigh the lower failure probability and oracle overhead of a few larger stages. For $s=n/4$ the overhead vanishes, but the reduction decreases slowly with $n$, because under the logarithmic-depth decomposition a register of $n/4$ qubits is shallower than the full register by only a near-constant depth. With $s=\lceil\sqrt{n}\rceil$ the depth of each partial diffusion operator instead approaches half that of the global one, as $\log s/\log n\to1/2$, and the reduction is retained.

The bottom row of Figure~\ref{fig:depth-scaling} shows the corresponding change in expected total CX gate count. While the MCX decomposition has a depth logarithmic in the number of control qubits, the number of CX gates in the transpiled circuit still grows linearly. The relative reductions in CX gate count are larger than those in depth, and for the stage sizes that grow polynomially with $n$ they no longer shrink towards zero with increasing $n$. The $s=\lceil\sqrt{n}\rceil$ strategy still uses fewer CX gates than $s=n/4$, from $n=56$, $72$ and $100$ at $r=1$, $5$ and $10$. At $n=256$ the CX count is reduced by $37.80\%$, $12.60\%$ and $6.87\%$ at $r=1$, $5$ and $10$ for $s=n/4$, and by $47.22\%$, $15.72\%$ and $8.56\%$ for $s=\lceil\sqrt{n}\rceil$. With $s=\lceil\log_2 n\rceil$ and $s=7$, the CX count at $n=256$ and $n=252$ is $38.24\%$ and $22.11\%$ lower than for Grover search at $r=1$, but $9.96\%$ and $39.13\%$ higher at $r=10$, as the oracle overhead again dominates once the oracle is costly.

Under this model, these results show that the recursive scheme reduces circuit depth relative to Grover search when the stage size grows with the problem size, and that, for stages of at least $\sqrt{n}$ qubits, a reduction persists when the oracle circuit is substantially deeper than the global diffusion operator. As an example, consider an oracle ten times as deep as the global diffusion operator, comparable to estimates for AES-128~\cite{implementation_efficient, cryptography2}. At $n=128$, four stages of $s=n/4$ qubits would have a failure probability of $9.31\times10^{-10}$ and an oracle overhead of $1.40\times10^{-7}\%$. They would reduce the expected depth of the diffusion operators by $33.14\%$, which gives an expected total depth reduction of $3.01\%$. Stages of $s=\lceil\sqrt{n}\rceil$ qubits, at the nearest plotted size above it, $n=132$, would have a failure probability of $2.68\times10^{-3}$ and an oracle overhead of $0.40\%$. They would reduce the expected depth of the diffusion operators by $56.63\%$, which gives an expected total depth reduction of $4.78\%$.

\section{Discussion and Future Work}

Our central contribution is establishing that the quadratic speedup of quantum search can be retained for search problems such as single-target unstructured search, even when only a small subset of qubits is operated on between any two successive oracle calls. We achieve this through a recursive construction whose dynamics admit an exact analysis when the state preparation unitary and the projector onto the marked subspace factorise across the same partition of the search register, as in Equation~\ref{eq:decomp}.  This has considerable practical and conceptual implications which we discuss below.

\paragraph{Recursive structure and eigenvalue degeneracy}
The key mathematical insight enabling our construction is the collapse of the principal angles between successive reflection eigenspaces to at most two distinct values. The eigenspaces $\mathcal{E}_W$ and $\mathcal{E_S}$ from Equation~\ref{eq:spaces} are each of dimension $2^{i-1}$ and one might expect the principal angles between them to span a correspondingly complex spectrum. That they reduce to a scalar $2\times2$ problem, as per Equation~\ref{eq:hermetian}, is an intriguing result and makes the analysis of the algorithm tractable. It is therefore natural to ask whether such degeneracies can arise within other quantum algorithms, such as quantum walks and phase estimation.

This tractability follows from defining the recursion such that each layer is itself a reflection, a structure that may generalise to other recursive quantum algorithms. By recursing through reflections, one obtains a well-defined rotation angle at each layer, analogous to Equation~\ref{equation:gamma_recursion}, and can precisely characterise how the target amplitude evolves through the intermediate stages, providing valuable insights into the optimal conditions for such approaches. Recursion in quantum search and amplitude amplification has a substantial prior literature, particularly within spatial search~\cite{scott_paper}, cryptanalysis~\cite{recursive_search}, and is central to the partial diffusion-based search algorithm of~\cite{hardwareGrover2}. However, as amplitude amplification with an unknown initial success probability is a bounded-error procedure~\cite{AmplitudeAmplification}, recursively calling it as a subroutine incurs a poly-logarithmic query overhead to guarantee a high probability of success~\cite{error_bound_1, scott_paper, recursive_search}. Prior work shows that this overhead can be avoided by carefully under-amplifying the intermediate stages and increasing the number of subdivisions as the recursion deepens. Adapting our work to other constructions can offer valuable insights into key challenges beyond search, and has potentially important implications for quantum algorithm design.

\paragraph{Practical implications}
Our evaluation in Section~\ref{sec:evaluation} shows that the theoretical reductions in non-oracle gate count carry over to transpiled circuits, in both gate count and circuit depth. Shallower circuits accumulate less gate error on noisy devices~\cite{benchmarking-and} and yield a reduction in circuit runtime that benefits quantum computing at any scale. On an 18-qubit search problem, partitioning into two stages reduces the expected diffusion circuit depth by as much as $40\%$--$76\%$, depending on the gate decomposition used, at a cost of only $0.4\%$ additional expected oracle calls. The three-stage configuration achieves even larger depth reductions ($50\%$--$86\%$) at a higher oracle overhead of $6.7\%$. Extending to larger search sizes, the failure probability and oracle overhead diminish rapidly with $n$ when the size of each partitioned register grows polynomially with $n$, and remain bounded when it is at least $\log_2 n$. When the oracle depth is equal to that of the global diffusion operator, partitioning the search into stages of $\lceil\sqrt{n}\rceil$ qubits reduces the expected total circuit depth by $28\%$ and the CX gate count by $47\%$ at $n=256$. When the oracle is ten times as deep, the depth reduction persists at approximately $5\%$.

All comparisons were performed on a fully connected topology, and the scaling analysis used a logarithmic-depth MCX decomposition, which favours Grover search. On connectivity-constrained hardware, controlled operations between distant qubits require significant SWAP routing to implement~\cite{routinng_1},  and with such routing overhead we would expect the depth reductions to be more pronounced. Furthermore, our construction is particularly well-suited for distributed architectures. When qubits are grouped into processing nodes with limited inter-node connectivity, all non-oracle operations can be executed entirely within a single node, as in S1, or in S3 when each node has its own ancillae, with only the oracle requiring communication between processors. Recent experiments on distributed Grover search~\cite{distributed_demonstration} have demonstrated that inter-node entanglement generation introduces substantially more noise than local operations --- precisely the overhead our approach eliminates from the diffusion steps.

The algorithm's flexibility in partition size and ordering further supports hardware-aware deployment. Partitioned registers do not need to be uniform and boundaries can be aligned with the target device, provided each register satisfies the tensor product constraint. This adaptability, combined with the quantifiable trade-off between partition count and oracle overhead established in Subsection~\ref{subsec:oracle}, allows the algorithm to be tuned to the specific connectivity and noise characteristics of the underlying hardware. That said, a detailed analysis of how noise propagates through the recursive construction would be essential in fully evaluating the approach on real hardware. In particular, understanding how gate errors accumulate in the recursively defined rotation angle $\gamma_i$ from Equation~\ref{equation:gamma_recursion} would be key to assessing performance on near-term devices.

\paragraph{The role of non-oracle operators in quantum search}
Perhaps the most notable conceptual implication is what our result reveals about the role of the non-oracle operators in quantum search. In standard amplitude amplification, both the oracle and the diffusion operator act on every qubit at every iteration, although for unstructured search the latter can be replaced by a single-qubit gate on each~\cite{kato2005single, jiang2017transverse}. Previous works~\cite{grover_trade-offs_2002,groverOptGates, hardwareGrover2} have shown that the non-oracle operators applied between successive oracle calls can be made sparse \textit{on average}, acting on only a few qubits at a time while retaining near-optimal query complexity, although each still applies a global reflection. The construction of~\cite{hardware} gives numerical evidence that, for unstructured search, the non-oracle operators can each act on as few as half of the qubits with a query complexity similar to that of Grover search. Our result establishes that the quadratic speedup of amplitude amplification can be retained even when only a small fraction of the qubits is acted on between any two successive oracle calls, as few as $\log_2(\log_2 N)$ for single-target unstructured search. More generally, for amplitude amplification in which the state preparation unitary and the marked-subspace projector factorise across the same partition of the search space, we can construct and analyse an algorithm in which \textit{every} non-oracle operator acts on only a single register of this partition. We show that this algorithm achieves the quadratic speedup when the sum of the local success probabilities remains bounded (Subsection~\ref{subsec:oracle}), with the oracle as the only operation acting across the partition.

We observe that, if the goal is simply to minimise the number of qubits acted on between successive oracle calls, this can also be done by padding existing algorithms with pairs of oracle calls that cancel to the identity. For any search algorithm that has both an $O(\sqrt{N})$ oracle complexity and an $O(\sqrt{N})$ non-oracle gate count, such as that of~\cite{hardwareGrover2} at fixed $\epsilon$, this would add a constant-factor overhead to the number of oracle calls. Inserting such a pair between every two consecutive non-oracle gates in the algorithm would leave either a single elementary gate, acting on only one or two qubits, or the identity between any two successive oracle calls.  However, while this meets the goal as stated, it does not amount to the same result. We can clarify the distinction by observing the effect each approach has on the information about the target that the oracle introduces through its phase flip. In Grover search this information allows the diffusion operator to concentrate amplitude from every basis state onto the target, so the information from each oracle reaches all $N$ basis states before the next. Our approach shows that this reach can be significantly restricted without losing the quadratic speedup. For single-target unstructured search, the information imparted by each oracle can reach as few as $\log_2 N$ basis states (a register of $\log_2(\log_2 N)$ qubits) before the next and retain this speedup, as there is always a single partial diffusion operator between successive oracle calls (Subsection~\ref{subsec:oracle}). In the padded algorithm, by contrast, this reach is not always restricted once we consider the information that the system actually retains. The information from the first oracle of each inserted pair is subsequently removed by the second. Thus only the original oracles contribute retained information. In the global reflection steps of the original algorithm, this information still reaches all $N$ basis states before the next contributing oracle. The padded algorithm therefore does not have the same effect as our construction, as it does not apply the same restriction to the reach of the information imparted by each oracle. However, this does raise the broader question of how many basis states the information from each oracle actually needs to reach before the next to retain a quantum advantage in search.

\paragraph{Limitations}
We outline several potential limitations of the approach and potential avenues to address them. Firstly, the ceiling on the success probability of a single run is below one, as it is bounded by the maximum of the final-register probability $P(x_m)$, which by Equation~\ref{eq:stage-prob} is less than one when $|\gamma_m|<\theta_m$. The success probability bound in Equation~\ref{eq:overall-succ} also includes a rounding error from the integer iteration requirement on the final register, in addition to the residual term $2\sin(\theta_m)\tan(2|\gamma_m|)$. This rounding error can be larger than in standard quantum search, as $|\gamma_m| \gg \theta$ in general. Altering the phase of the partial diffusion operators may eliminate this error entirely, analogous to existing proposals for standard quantum search~\cite{AbitraryPhases, DeterministicRestrictedOracle, DeterministicAdjustableParameters}. Combined with further iterations at the penultimate stage, it could also remove the ceiling on $P(x_m)$ by bringing $|\gamma_m|$ to $\theta_m$ (Equation~\ref{equation:gamma_recursion}), although fully deterministic variants would also require exact recovery of the lower registers. Without such variants, several runs may be necessary when the partitions are small and the target success probability is high. For the 18-qubit search with a 3-stage configuration, reaching a minimum success probability of $0.99$ requires up to two runs and $86\%$ more oracle calls than Grover search on the same problem (Subsection~\ref{subsec:sim-depth}). Furthermore, our analysis guarantees that the expected oracle overhead relative to amplitude amplification remains constant as the problem size grows only when the sum of the local probabilities $\sum_i p_i=\sum_i\sin^2(\theta_i)$ remains bounded (Subsection~\ref{subsec:oracle}). For unstructured search with a single target, this requires the stage size to grow at least logarithmically with $n$.

Secondly, choosing the iteration count $t_m^*$ requires knowledge of the local success probabilities $p_i$, which are known in single-target unstructured search but may not be in general. Over-amplification is a related concern, as exceeding the optimal iteration count causes the target amplitude to decrease. Fixed-point variants of amplitude amplification~\cite{FixedPointSearch, fixed-point} may address this, and so relax the need to know the $p_i$, while preserving the quadratic speedup, and their adaptation is a natural direction for future work.

Finally, the tensor product decomposition of both the state preparation unitary and the marked-subspace projector across the chosen registers is central to our analysis. For the state preparation operator, this constraint enables the construction of the partial diffusion operators: $S_{\psi_i}$ requires the ability to reflect about $\ket{\psi_i}$ on register $i$ independently, which the factorisation $A = \bigotimes_i A_i$ guarantees. Unstructured search satisfies this naturally, but for algorithms that generate inter-register entanglement during state preparation, the approach cannot be directly applied. Approximate tensor decompositions of the state preparation unitary --- for instance via low-rank operator-Schmidt truncations~\cite{tensor_decomp} --- may enable extensions of the approach to such settings, although such truncations do not always yield unitary operators. For the marked subspace, the factorisation $P_G = \bigotimes_i P_{G_i}$ enables the eigenvalue reduction central to the analysis in Subsection~\ref{sec:proof}. Without this factorisation, preservation of the tensor product subspace $\mathcal{S}_m$ and the reduction to two distinct principal angles are no longer guaranteed. The operators remain well-defined, but the analysis may need to account for multiple independent rotation angles in place of the single recursively defined angle $\gamma_i$. Extending the analysis to marked subspaces that do not factorise across the chosen registers is a promising direction for future work.

\section{Conclusion}

We have shown that the quadratic speedup of quantum search can be retained for search problems such as single-target unstructured search, even when only a small subset of qubits is operated on between any two successive oracle calls. We achieve this through a novel recursive construction of reflections using only the oracle and local diffusion operators. When the state preparation unitary and the projector onto the marked subspace factorise across the same partition of the search space --- such that $A = \bigotimes_{i=1}^{m}A_i$ and $P_G = \bigotimes_{i=1}^{m}P_{G_i}$ --- the construction admits an exact expression for its dynamics. This is enabled by an intriguing degeneracy in principal angles between the applied operators, which collapse to at most two distinct values and are governed by a single recursively defined angle determined by the local overlaps between the initial and target states. The approach yields an oracle overhead per run of $O\!\left(1\big/\prod_{i=1}^{m}\cos(\theta_i)\right)$ relative to amplitude amplification, where $m$ is the number of partitioned registers and $p_i=\sin^2(\theta_i)$ is the initial local success probability in register $i$. For $0<p_i<1/2$, this overhead, and its expected value over repeated runs, remains $O(1)$ as the initial success probability $p\to0$ whenever $\sum_{i=1}^{m}p_i=O(1)$.

We apply our approach to single-target unstructured search, a problem that naturally satisfies both factorisation assumptions, and show that our approach retains the $O(\sqrt{N})$ oracle complexity when each partition contains at least $\log_2(\log_2 N)$ qubits. On an 18-qubit search problem, partitioning into three stages reduces the expected diffusion circuit depth by as much as $50\%$--$86\%$ compared to Grover, depending on the gate decomposition used, requiring $6.7\%$ additional expected oracle calls. For larger search sizes, with stages of at least $\sqrt{n}$ qubits, the oracle overhead and failure rate diminish rapidly with problem size. Valuable reductions in circuit depth and CX gate count then persist even when the oracle circuit is substantially deeper than the diffusion operator. The algorithm offers significant flexibility in choosing register partitions that can be tailored to the hardware topology and is particularly well suited to distributed architectures where all non-oracle operations can be confined to individual processors.

More broadly, our results show that quantum search does not need to act on more than a small subset of qubits between successive oracle calls to achieve a quadratic speedup. Furthermore, the structural degeneracy that reduces the analysis of our algorithm to a scalar $2 \times 2$ problem may exist in other quantum algorithms, potentially enabling them to be constructed without global operators while preserving their computational advantage.

\section*{Code Availability}
The Qiskit implementation of the recursive search algorithm and supporting code are available at~\cite{burke2026code}.
\printbibliography
\appendix

\section{Analysing the lower registers}
\label{app:run}
In this appendix we derive the state of the intermediate registers after applying $(S_{\psi_m}W_{m-1})^{t_m}$ to the initial state $\ket{\psi}$. We use this to show how applying $(S_{\psi_k}W_{k-1})^{-t_k}=(W_{k-1}S_{\psi_k})^{t_k}$ at each remaining stage $k=m-1,\ldots,1$ can increase the probability of obtaining the corresponding local target $\ket{x_k}$. Here $t_k$ is the same iteration count used to construct $W_k$ within the recursive construction of $W_{m-1}$. This allows the full target to be recovered with an exactly determined success probability. Throughout, we assume at least two stages ($m \geq 2$), integer iteration counts $t_i \geq 1$ for $i < m$, $t_m \geq 0$, $0<\theta_i<\pi/4$ for every register, and $\sin(2t_i\gamma_i)\neq0$ for $i<m$, so that $0<|\gamma_k|<\pi/4$ for $k \leq m$ (Subsection~\ref{sec:proof}). As in Subsection~\ref{sec:proof}, we analyse the component in which registers $m,\ldots,k+1$ are in $\ket{x_m,\ldots,x_{k+1}}$, restricting the operators to $\mathcal{S}_k$. The projected states below are unnormalised unless stated otherwise. For even $m$, removing the final oracle call of the run (Subsection~\ref{subsec:oracle}) leaves Equations~\ref{eq:app-full} and~\ref{eq:app-prefix} unchanged, as it immediately precedes the measurement and commutes with the projectors onto the target components.

We first derive the action of $t\geq0$ iterates, or of their inverse, on $\ket{\psi_k,\ldots,\psi_1}$ for $2\leq k\leq m$, after projecting register $k$ onto $\ket{x_k}$. Recall from Subsection~\ref{sec:proof} that
\begin{equation}
    Q_k=(S_{\psi_k}W_{k-1})^{t_k}P_{\psi_k}(W_{k-1}S_{\psi_k})^{t_k}.
\end{equation}
As in Subsections~\ref{sec:proof} and~\ref{sec:succ}, let $\ket{\phi_l^{(1)}}$ and $\ket{\phi_l^{(2)}}$ denote the eigenvectors of $T$ (Equation~\ref{eq:T}, with $i=k$) with eigenvalues $\cos^2(\gamma_k)$ and $\sin^2(\gamma_k)$, respectively. The initial state of the lower registers decomposes in this basis as
\begin{equation}
    \ket{\psi_{k-1},\ldots,\psi_1}
    = \sum_l\bigl(\alpha_l\ket{\phi_l^{(1)}}+\beta_l\ket{\phi_l^{(2)}}\bigr).
\end{equation}
The operator $(2T-\mathbb{I})/\cos(2\gamma_k)$ acts as $+1$ on the first set of eigenvectors and $-1$ on the second. Applying it to this decomposition therefore gives
\begin{equation}
    \frac{2T-\mathbb{I}}{\cos(2\gamma_k)}\ket{\psi_{k-1},\ldots,\psi_1}
    = \sum_l\bigl(\alpha_l\ket{\phi_l^{(1)}}-\beta_l\ket{\phi_l^{(2)}}\bigr).
\end{equation}
Since $\ket{\psi_k,\ldots,\psi_1}=\ket{\psi_k}\ket{\psi_{k-1},\ldots,\psi_1}$, this decomposition also gives
\begin{equation}
    \ket{\psi_k,\ldots,\psi_1}
    =\sum_l\bigl(\alpha_l\ket{u_l^{(1)}}+\beta_l\ket{u_l^{(2)}}\bigr),
\end{equation}
where $\ket{u_l^{(j)}}=\ket{\psi_k}\ket{\phi_l^{(j)}}$. The iterate rotates the planes corresponding to the two sets by $2\gamma_k$ and $\pi-2\gamma_k$, respectively, and its inverse reverses both rotations. Using the expressions for $\ket{w_l}$ in Subsections~\ref{sec:proof} and~\ref{sec:succ}, together with
\begin{equation*}
    \frac{\sin(2\theta_k)}{\sin(2\gamma_k)}
    =\frac{1}{\sin(2t_{k-1}\gamma_{k-1})},
\end{equation*}
the projection onto $\ket{x_k}$ after $t$ forward or inverse iterates is
\begin{equation}
\begin{split}
    (\bra{x_k}\otimes\mathbb{I})&(S_{\psi_k}W_{k-1})^{\pm t}\ket{\psi_k,\ldots,\psi_1}\\
    ={}&\sin(\theta_k)\cos(2t\gamma_k)
    \sum_l\bigl(\alpha_l\ket{\phi_l^{(1)}}
    +(-1)^t\beta_l\ket{\phi_l^{(2)}}\bigr)\\
    &\mp\frac{\cos(\theta_k)\sin(2t\gamma_k)}
    {\sin(2t_{k-1}\gamma_{k-1})}(Q_{k-1}-P_\psi)
    \sum_l\bigl(\alpha_l\ket{\phi_l^{(1)}}
    -(-1)^t\beta_l\ket{\phi_l^{(2)}}\bigr).
\end{split}
\label{eq:app-iter}
\end{equation}
The upper sign corresponds to forward iterates and the lower sign to inverse iterates.

Using $P_\psi\ket{\psi_{k-1},\ldots,\psi_1}=\ket{\psi_{k-1},\ldots,\psi_1}$, $Q_{k-1}^2=Q_{k-1}$, and Equation~\ref{equation:Q_diag_psi}, together with $T=\sin^2(\theta_k)Q_{k-1}+\cos^2(\theta_k)P_\psi$, gives
\begin{equation}
\begin{aligned}
    (2T-\mathbb{I})\ket{\psi_{k-1},\ldots,\psi_1}
    &=
    \ket{\psi_{k-1},\ldots,\psi_1}
    -2\sin^2(\theta_k)(\mathbb{I}-Q_{k-1})
    \ket{\psi_{k-1},\ldots,\psi_1},\\
    (Q_{k-1}-P_\psi)\ket{\psi_{k-1},\ldots,\psi_1}
    &=
    -(\mathbb{I}-Q_{k-1})\ket{\psi_{k-1},\ldots,\psi_1},\\
    (Q_{k-1}-P_\psi)(\mathbb{I}-Q_{k-1})
    \ket{\psi_{k-1},\ldots,\psi_1}
    &=
    -\sin^2(2t_{k-1}\gamma_{k-1})
    \ket{\psi_{k-1},\ldots,\psi_1}.
\end{aligned}
\end{equation}
These identities express the projected state in Equation~\ref{eq:app-iter} as a linear combination of $\ket{\psi_{k-1},\ldots,\psi_1}$ and $(\mathbb{I}-Q_{k-1})\ket{\psi_{k-1},\ldots,\psi_1}$.

For the final stage, we take $k=m$, $t=t_m$ and the upper sign in Equation~\ref{eq:app-iter}. Writing $\ket{v}=(S_{\psi_m}W_{m-1})^{t_m}\ket{\psi}$ and substituting the identities above gives
\begin{equation}
\begin{split}
    (\bra{x_m}\otimes\mathbb{I})\ket{v}
    ={}&\mu_{m-1}(\mathbb{I}-Q_{m-1})\ket{\psi_{m-1},\ldots,\psi_1}\\
    &+\nu_{m-1}\ket{\psi_{m-1},\ldots,\psi_1},
\end{split}
\end{equation}
with real coefficients
\begin{equation}
\begin{aligned}
    t_m\text{ even:}\quad
    &\mu_{m-1} =
    \frac{\cos(\theta_m)\sin(2t_m\gamma_m)}
    {\sin(2t_{m-1}\gamma_{m-1})\cos(2\gamma_m)},\\
    &\nu_{m-1} =
    \frac{\sin(\theta_m)\cos(2(t_m+1)\gamma_m)}
    {\cos(2\gamma_m)},\\[4pt]
    t_m\text{ odd:}\quad
    &\mu_{m-1} =
    \frac{\cos(\theta_m)\sin(2t_m\gamma_m)}
    {\sin(2t_{m-1}\gamma_{m-1})}
    -\frac{2\sin^3(\theta_m)\cos(2t_m\gamma_m)}
    {\cos(2\gamma_m)},\\
    &\nu_{m-1} =
    \frac{\sin(\theta_m)\cos(2t_m\gamma_m)}
    {\cos(2\gamma_m)}.
\end{aligned}
\label{eq:end-coef}
\end{equation}

We now apply $t_k$ inverse iterates at each stage $k=m-1,\ldots,1$. These preserve the target subspace of the higher registers, allowing us to analyse the projected lower-register state
\begin{equation}
    \mu_k(\mathbb{I}-Q_k)\ket{\psi_k,\ldots,\psi_1}
    +\nu_k\ket{\psi_k,\ldots,\psi_1}.
\end{equation}
The definition of $Q_k$ gives
\begin{equation}
    (W_{k-1}S_{\psi_k})^{t_k}(\mathbb{I}-Q_k)
    =(\mathbb{I}-P_{\psi_k})(W_{k-1}S_{\psi_k})^{t_k}.
\end{equation}
Using this identity and Equation~\ref{eq:app-iter} with the lower sign and $t=t_k$, the component with coefficient $\mu_k$ maps to the second term of Equation~\ref{eq:app-iter}, and applying the inverse iterates and projecting onto $\ket{x_k}$ leaves a state of the same form on the remaining registers, with coefficients
\begin{equation}
\begin{aligned}
    t_k\text{ odd:}\quad
    &\mu_{k-1}
    =-\frac{\cos(\theta_k)\sin(2t_k\gamma_k)}
    {\sin(2t_{k-1}\gamma_{k-1})}(\mu_k+\nu_k)
    -\frac{2\sin^3(\theta_k)\cos(2t_k\gamma_k)}
    {\cos(2\gamma_k)}\nu_k,\\
    &\nu_{k-1}
    =\frac{\sin(\theta_k)\cos(2t_k\gamma_k)}
    {\cos(2\gamma_k)}\nu_k,\\[4pt]
    t_k\text{ even:}\quad
    &\mu_{k-1}
    =-\frac{\cos(\theta_k)\sin(2t_k\gamma_k)}
    {\sin(2t_{k-1}\gamma_{k-1})\cos(2\gamma_k)}(\mu_k+\nu_k),\\
    &\nu_{k-1}
    =\sin(\theta_k)\bigl[
    \cos(2t_k\gamma_k)\nu_k
    +\tan(2\gamma_k)\sin(2t_k\gamma_k)(\mu_k+\nu_k)
    \bigr],
    \qquad k\geq2.
\end{aligned}
\label{eq:app-rec}
\end{equation}

At $k=1$, using
\begin{equation*}
    (S_{\psi_1}W_0)^{t_1}\ket{\psi_1}
    =(-1)^{t_1}\bigl[
    \cos(2t_1\theta_1)\ket{\psi_1}
    +\sin(2t_1\theta_1)\ket{\psi_1^\perp}
    \bigr]
\end{equation*}
from Subsection~\ref{sec:proof}, the inverse iterates give
\begin{equation}
\begin{aligned}
    \bra{x_1}(W_0S_{\psi_1})^{t_1}(\mathbb{I}-Q_1)\ket{\psi_1}
    &=(-1)^{t_1+1}\sin(2t_1\theta_1)\cos(\theta_1),\\
    \bra{x_1}(W_0S_{\psi_1})^{t_1}\ket{\psi_1}
    &=(-1)^{t_1+1}\sin((2t_1-1)\theta_1).
\end{aligned}
\label{eq:app-base}
\end{equation}
The final projection therefore gives the full target success probability
\begin{equation}
    P^*(x)=\bigl[
    \sin(2t_1\theta_1)\cos(\theta_1)\mu_1
    +\sin((2t_1-1)\theta_1)\nu_1
    \bigr]^2.
\label{eq:app-full}
\end{equation}

The probability that registers $m,\ldots,k+1$ yield target outcomes, irrespective of the remaining outcomes, is the squared norm of the projected state. This norm is preserved by the subsequent inverse iterates. Equation~\ref{equation:Q_diag_psi}, or direct calculation for $k=1$, gives both the squared norm of $(\mathbb{I}-Q_k)\ket{\psi_k,\ldots,\psi_1}$ and its inner product with $\ket{\psi_k,\ldots,\psi_1}$ as $\sin^2(2t_k\gamma_k)$. Hence,
\begin{equation}
    \Pr(x_m,\ldots,x_{k+1})
    =(\mu_k^2+2\mu_k\nu_k)\sin^2(2t_k\gamma_k)+\nu_k^2,
    \qquad 1\leq k\leq m-1.
    \label{eq:app-prefix}
\end{equation}
Together, these expressions give the exact probabilities of recovering the full target or its higher-register components under the stated assumptions. For odd $t_k$ with $2\leq k<m$, Equation~\ref{eq:app-rec} preserves $\nu_k=0$, allowing us to follow the component with coefficient $\mu_k$ separately and bound the contribution of the initial residual. Even counts can generate additional contributions proportional to the initial state of the remaining registers, whose interference can increase or decrease the success probability. We therefore derive the following bound for odd intermediate counts, while allowing either parity for $t_1$ and $t_m$.

We first bound the initial residual coefficient $\nu_{m-1}$ when $t_m=t_m^*$. Since $t_m^*$ is the integer nearest $\pi/(4|\gamma_m|)-1/2$, rounding gives
\begin{equation*}
    \left|(2t_m+1)|\gamma_m|-\frac{\pi}{2}\right|
    \leq|\gamma_m|.
\end{equation*}
Hence both $|\cos(2t_m\gamma_m)|$ and $|\cos(2(t_m+1)\gamma_m)|$ are at most $\sin(2|\gamma_m|)$. Substituting into Equation~\ref{eq:end-coef} gives
\begin{equation}
    |\nu_{m-1}|\leq\sin(\theta_m)\tan(2|\gamma_m|).
\end{equation}

Following the component with coefficient $\mu_{m-1}$ alone, Equation~\ref{eq:app-rec} gives $\nu_k=0$ at every subsequent stage. The sine factors cancel between successive stages and the final projection at $k=1$, leaving a contribution to the full target amplitude of magnitude
\begin{equation*}
    |\mu_{m-1}|\,|\sin(2t_{m-1}\gamma_{m-1})|
    \prod_{k=1}^{m-1}\cos(\theta_k).
\end{equation*}
The projected state after the final-stage amplification has norm $\sqrt{P(x_m)}$. Removing its residual component, whose norm is $|\nu_{m-1}|$, changes this norm by at most $|\nu_{m-1}|$. The remaining component therefore satisfies
\begin{equation*}
    \Bigl|\sqrt{P(x_m)}
    -|\mu_{m-1}|\,|\sin(2t_{m-1}\gamma_{m-1})|\Bigr|
    \leq|\nu_{m-1}|.
\end{equation*}
Combining this with the amplitude contribution above and the residual contribution of at most $|\nu_{m-1}|$, since the inverse iterates and target projections do not increase its norm, gives
\begin{equation}
\begin{split}
    \Bigl|\sqrt{P^*(x)}
    -\sqrt{P(x_m)}\prod_{k=1}^{m-1}\cos(\theta_k)\Bigr|
    &\leq\left(1+\prod_{k=1}^{m-1}\cos(\theta_k)\right)|\nu_{m-1}|\\
    &\leq2|\nu_{m-1}|\\
    &\leq2\sin(\theta_m)\tan(2|\gamma_m|).
\end{split}
\end{equation}

Thus, the full target success probability is approximately the final-register success probability multiplied by $\prod_{k=1}^{m-1}\cos^2(\theta_k)$, with an error that vanishes as $|\gamma_m|\to0$ for odd intermediate counts and $t_m=t_m^*$.

\end{document}